\documentclass[preprint,12pt]{elsarticle}

\usepackage{amssymb}
\usepackage{amsmath}
\usepackage{graphicx}
\usepackage{graphics}
\usepackage{multirow}
\usepackage{comment}
\usepackage{hyperref}
\usepackage{enumitem}
\usepackage{makecell}
\usepackage{graphicx}
\usepackage[ruled,vlined]{algorithm2e}
\usepackage{xcolor}

\usepackage{tabularray}
\usepackage{float}
\usepackage{subfigure}
\usepackage{subcaption}
\usepackage{ifthen}
\usepackage{amsthm}
\usepackage{bbm}
\usepackage{tcolorbox}
\usepackage{booktabs}
\usepackage{tabularx}
\usepackage{array}
\newboolean{showcomments}
\setboolean{showcomments}{true}
\ifthenelse{\boolean{showcomments}}
{\newcommand{\nb}[2]{
		\fcolorbox{black}{yellow}{\bfseries\sffamily\scriptsize#1}
		{\sf\small$\blacktriangleright$\textit{#2}$\blacktriangleleft$}
	}
	
}
{\newcommand{\nb}[2]{}
	
}

\journal{}

\begin{document}

\begin{frontmatter}

\title{Statistical-Based Metric Threshold Setting Method for Software Fault Prediction in Firmware Projects: An Industrial Experience}

\author[label1]{Domenico Amalfitano}
\author[label1]{Marco De Luca}
\author[label1]{Anna Rita Fasolino}
\author[label1]{Porfirio Tramontana}

\affiliation[label1]{organization={Department of Electrical Engineering and Information Technology (DIETI), University of Naples "Federico II" },
            addressline={Via Claudio 21}, 
            city={Naples},
            postcode={80125}, 
            state={Italy}
            }

\begin{abstract}

Ensuring software quality in embedded firmware is critical, especially in safety-critical domains such as automotive systems, where compliance with functional safety standards like ISO 26262 requires strong guarantees of software reliability. While machine learning-based fault prediction models have demonstrated high accuracy, their lack of transparency and interpretability limits their adoption in industrial settings. Developers need actionable insights that can be directly employed in software quality assurance (SQA) processes and guide defect mitigation strategies. 
In this paper, we present a structured process for defining context-specific software metric thresholds suitable for integration into fault detection workflows in industrial settings. Our approach supports cross-project fault prediction by deriving thresholds from one set of projects and applying them to independently developed firmware, thereby enabling reuse across similar software systems without retraining or domain-specific tuning.
We analyze three real-world C-embedded firmware projects provided by an industrial partner, using Coverity and Understand static analysis tools to extract software metrics. Through statistical analysis and hypothesis testing, we identify discriminative metrics and derived empirical threshold values capable of distinguishing faulty from non-faulty functions. The derived thresholds are then validated through an experimental evaluation, demonstrating their effectiveness in identifying fault-prone functions with high precision. The results confirm that statistically derived thresholds can serve as a practical and interpretable solution for fault prediction, aligning with industry standards and SQA practices. This approach provides a practical alternative to black-box AI models, allowing developers to systematically assess software quality, take preventive actions, and integrate metric-based fault prediction into industrial development workflows to mitigate software faults.
\end{abstract}


\end{frontmatter}

\section{Introduction}
\label{sec:introduction}

Ensuring software quality is a fundamental challenge in embedded firmware development, especially within safety-critical industries like automotive engineering. Compliance with stringent safety standards such as ISO 26262 \cite{iso26262-1-2018} requires maintaining rigorous software quality to prevent faults and ensure system reliability \cite{Ardila17, Bahig17, Hocking14, Santilli24}. In recent years, several studies have highlighted the importance of integrating product and process metrics into software development workflows to support compliance with ISO 26262 and enhance safety assurance activities in industrial contexts \cite{Hillenbrand2011, Luo2015,Luo2016, Tabani2019, Maurer2014}. In particular, the study by Vogel \textit{et al}. \cite{Vogel2021} highlights that although numerous metrics aligned with standards such as ISO 26262 are available, their practical application in industry remains limited due to a lack of contextual clarity and integration strategies. As also emphasized by Mori \textit{et al.}~\cite{mori2018}, the absence of domain-independent thresholds significantly limits the practical applicability of software metrics, as it prevents their straightforward reuse across different projects and domains. Without empirically validated thresholds that can generalize beyond specific case studies, organizations are often forced to redefine acceptable metric ranges from scratch, reducing consistency and increasing the cost and effort required to integrate metric-based assessments into industrial quality assurance workflows. This underscores the need for empirically grounded methodologies that define how metrics and thresholds can be tailored to the specific characteristics of industrial software projects. These works emphasize that metric-driven approaches not only help in evaluating software artifacts and guiding architectural refinements but also offer quantitative evidence of compliance, thereby reducing the cost and effort associated with safety certification and assessment.
Among various strategies to achieve this compliance, Software Quality Assurance (SQA) processes are widely adopted by industries to systematically identify and address software faults early in the development lifecycle. In this context, software fault prediction plays a crucial role by highlighting fault-prone software components, enabling targeted quality assurance efforts, and optimizing testing resources. Effective fault prediction can significantly reduce maintenance costs, enhance reliability, and prevent software failures, thereby improving overall software quality \cite{Khoshgoftaar03, KHATRI23, Khan20}. However, developing a reliable fault prediction model requires accurate datasets, discriminative software metrics, and well-defined threshold values capable of differentiating between faulty and non-faulty functions.

Existing fault prediction approaches generally fall into two categories: statistical techniques and machine learning (ML)-based methods \cite{Rathore20171}. Statistical approaches rely primarily on historical defect data and code metrics, offering high interpretability and low computational complexity. However, they often struggle to capture complex or non-linear relationships within data. On the other hand, ML-based approaches have gained popularity due to their high predictive accuracy and ability to analyze complex, high-dimensional data \cite{Akimova2021}. Despite these strengths, ML models are not broadly used in industrial environments due to their limited transparency and interpretability \cite{Haldar23, Pandey2021}. Specifically, ML techniques often behave as \textit{black boxes}, producing predictions without clear explanations or actionable guidance on how developers can modify code to reduce fault-proneness. Although recent developments in explainable AI have attempted to address these interpretability challenges, the provided explanations typically remain too complex to be directly useful for practical code improvements by developers \cite{Begum2023, Santos2022}. Due to these limitations, industrial practitioners prefer metrics-based approaches, particularly those derived from static code analysis. Metrics-based methods offer clear and interpretable feedback, explicitly identifying problematic areas in the code and guiding developers toward actionable improvements \cite{Filo24, Sultan2021, FERREIRA2012}. However, metrics alone provide insufficient guidance unless coupled with clearly defined threshold values. Thresholds distinguish between acceptable and problematic values of metrics, enabling developers to easily recognize when code adjustments are necessary \cite{Mishra2021}. A major challenge in metrics-based fault prediction is the lack of universally applicable threshold values. Industries differ widely in their coding styles, programming languages, and application domains, making generalized metrics and thresholds impractical. Metrics and thresholds effective in one context rarely transfer directly to another, emphasizing the need to empirically derive context-specific threshold values tailored to each industry's unique software characteristics \cite{Zhang13, mori2018, Sultan2021, Santos17, Alves10}.

In this paper, we tackle the challenge of defining software metric thresholds suitable for integration into cross-project fault prediction processes in industrial settings. Our work focuses on designing and implementing a process tailored to our industrial partner's automotive embedded firmware development. The proposed approach includes the construction of a representative function-level dataset, the identification of discriminative software metrics, and the empirical derivation and validation of context-specific threshold values. 

The core contribution of this paper does not lie in proposing universally optimal threshold values, but rather in the structured process we designed to derive \textit{context-specific thresholds} and demonstrate their practical applicability in an industrial setting. This aligns with the challenge highlighted by Mori \textit{et al.}~\cite{mori2018}, who observe that the absence of domain-independent thresholds hampers the practical adoption of software metrics.

The main contributions are summarized in the following:
\begin{itemize}
    \item We propose a practical and replicable process for deriving software metric thresholds aimed at supporting fault prediction in industrial embedded software.
    \item {We apply this process to real-world firmware projects developed in the automotive domain, using metrics extracted from static analysis tools and fault data derived from issue tracking systems.}
    \item We validate the thresholds on an independent industrial project, demonstrating their effectiveness in identifying fault-prone functions with high precision.
\end{itemize}

By grounding our methodology in a real-world industrial setting, this work provides practitioners with a structured and well-defined fault prediction process that can be adopted or adapted to their specific development environments, enhancing software quality and supporting defect prevention.

The remainder of the paper is organized as follows. Section \ref{sec:related} reviews related works on existing fault prediction techniques. Section \ref{sec:indContext} describes our industrial context, including firmware architecture and development processes. Section \ref{sec:method_overview} presents an overview of the followed process. Section \ref{sec:dataCollection} details our dataset construction approach, while Section \ref{sec:prediction} discusses the selection of software metrics and derivation of threshold values. Section~\ref{sec:empEvaluation} presents the results of our experimental evaluation, while Section~\ref{sec:furdiscussion} provides a detailed discussion and interpretation of these findings. Section \ref{sec:Threats} presents the threats affecting the validity of this study.
Finally, Section \ref{sec:ConcFut} concludes the paper and outlines directions for future research.

\section{Related Works}
\label{sec:related}

Monitoring software metrics plays a crucial role in evaluating key software attributes such as fault-proneness, maintainability, reusability, and portability. In software engineering, fault prediction aims to identify modules likely to contain faults, enabling more efficient testing, improved software quality, and enhanced reliability. Fault prediction typically relies on analyzing code complexity, historical fault data, and developer activity to anticipate defects before they manifest in production \cite{BIGONHA2019}. Accurate fault prediction models contribute to reducing maintenance costs, optimizing resource allocation, and minimizing software failures, which ultimately enhance the overall user experience. Over the years, various fault prediction techniques have been proposed, which can be broadly classified into statistical methods and machine learning (ML)-based approaches \cite{Rathore20171}.

Despite the availability of numerous software metrics, their usefulness is limited unless they are associated with well-defined threshold values \cite{Filo24, Sultan2021, FERREIRA2012}. Thresholds provide actionable insights by distinguishing between acceptable and problematic metric values, allowing developers to assess software quality more effectively. According to Mishra \textit{et al.} \cite{Mishra2021}, threshold computation methods can be broadly categorized into three approaches: (i) expert-defined thresholds based on programmer experience, (ii) statistical thresholds derived from empirical data, and (iii) quality-related models that incorporate fault information. The approach adopted in this study falls within the quality-related model category, as we derive threshold values by analyzing fault-prone functions within real-world software projects.

\subsection{Dataset Building and Fault Prediction Granularity}

A fault prediction model relies heavily on the dataset construction process, as it determines the model's accuracy and applicability. According to the taxonomy of Rathore \cite{Rathore20171} \textit{et al.}, fault prediction can be performed using training and testing datasets from three distinct settings:

\begin{itemize}
    \item \textit{Intra-Release Prediction}: The dataset is split into training and testing subsets within the same software release, employing n-fold cross-validation.
    \item \textit{Inter-Release Prediction}: A fault prediction model is trained on earlier releases and tested on a later release of the same software project.
    \item \textit{Cross-Project Prediction}:  The model is trained on defect data from one or more projects and then used to predict defects in a different project, usually when historical defect data are not available. 
\end{itemize}

The dataset structure also varies based on the granularity of prediction, which defines the unit of analysis:

\begin{itemize}
    \item \textit{Binary Classification}: The most common approach, where software modules are labelled as either faulty (1) or non-faulty (0), yielding higher accuracy and recall values.
    \item \textit{Fault Count Prediction}: The dataset includes the number of times a module has triggered a fault.
    \item \textit{Severity-Based Prediction}: The dataset specifies both the frequency and severity of faults. However, defining fault severity is subjective and varies across organizations, making it difficult to standardize.
\end{itemize}

Prediction can be conducted at various code granularities, such as component, package, file, or function level. Function-level prediction, as adopted in this study, offers finer-grained insights into fault-proneness.

\subsection{Statistical Techniques}

Statistical methods for fault prediction rely on historical defect data and statistical measures to identify software components that are likely to contain faults. These techniques typically use regression models, correlation analysis, and hypothesis testing to establish relationships between software metrics and defect-proneness. One of the most widely used approaches is regression modeling, which predicts the number of faults based on past defect occurrences and various code-related attributes, such as object-oriented (OO) metrics and lines of code (LOC) \cite{Taskeen23}.

Rathore \textit{et al.} \cite{Rathore2012_2} applied logistic regression to assess the predictive power of OO metrics in identifying software defects. To mitigate redundancy and high correlations among metrics, Spearman's correlation analysis was performed, refining the metric set. The study concluded with a multivariate linear regression model, identifying the most influential metrics for fault prediction. Similarly, Elish \textit{et al.} \cite{ELISH2011} analyzed the Eclipse project, using correlation analysis to examine the relationship between package-level metrics and defect occurrence. Their findings demonstrated that the Martin Metric Suite is highly effective in predicting both pre-release and post-release faults in software packages.

Beyond regression-based models, hypothesis testing techniques have been employed to validate the relationship between software metrics and fault-proneness. Matsumoto \textit{et al.} \cite{Matsumoto2010} analyzed the impact of developer-related metrics, such as the number of contributors to a module, using hypothesis testing. Their study revealed that developer activity metrics are strong predictors of potential faults in software modules. Similarly, Rathore \textit{et al.} \cite{Rathore2012} investigated the predictive ability of OO metrics related to coupling, cohesion, inheritance, and complexity. Their findings indicate that while inheritance metrics alone do not strongly predict fault-proneness, their accuracy improves significantly when combined with coupling and complexity metrics.

\subsection{Learning Techniques}

Machine learning (ML) and deep learning (DL) techniques have significantly advanced software fault prediction by enabling models to automatically identify complex patterns from large datasets. These techniques have proven highly effective in detecting faulty software components, surpassing traditional statistical methods in handling high-dimensional data and intricate relationships. Popular approaches include decision trees, support vector machines (SVM), and neural networks, with deep learning models such as recurrent neural networks (RNNs) and convolutional neural networks (CNNs) gaining traction in fault prediction tasks \cite{Akimova2021}.

However, these models face two significant challenges: class imbalance and lack of interpretability, which may hinder their applicability and efficiency in predicting fault \cite{Jing17, Gezici2022}. The class imbalance problem arises when the dataset contains significantly more non-faulty instances than faulty ones. This imbalance skews model training, leading to biased predictions where the model tends to classify most instances as non-faulty, failing to adequately recognize fault-prone functions.
In addition to class imbalance, these models often operate as “black boxes", making it difficult to understand how they generate predictions. While these exhibit high predictive accuracy, their lack of interpretability raises concerns in critical domains where explainability is essential \cite{Haldar23}. Without a clear understanding of why a function is classified as faulty, developers may struggle to trust and act upon the model's predictions, limiting its usefulness in real-world software quality assurance processes.

To improve the interpretability of ML-based fault prediction, researchers have explored explainable AI (XAI) techniques. Esteves \textit{et al.} \cite{Esteves2020} applied SHapley Additive exPlanation (SHAP) to identify which features influenced defect predictions in a tree boosting algorithm (XGBoost), providing insights into model decision-making. Similarly, Begum \textit{et al.} \cite{Begum2023} utilized Local Interpretable Model-agnostic Explanations (LIME) to highlight the most significant features affecting model predictions, enhancing transparency and usability.  More recently, Santos \textit{et al.} \cite{Santos2022} conducted a large-scale analysis of 47,618 classes from 53 open-source Java projects, exploring how software features impact defect prediction. Their study identified key software attributes that correlate with defect-proneness, reinforcing the importance of interpretability in defect prediction models. These contributions demonstrate the growing need for explainable models that not only provide accurate predictions but also offer actionable insights for developers in managing software quality.

\subsection{Threshold Derivation Techniques}
\label{sec:Threshold Derivation Techniques}

Several approaches have been proposed in the literature to define thresholds for software metrics, each with its own assumptions and goals. The absence of standardized or universally applicable thresholds has led researchers to explore diverse strategies, often influenced by the nature of the dataset or the quality attribute of interest (e.g., fault-proneness, code smells, maintainability).

In their work, Alves \textit{et al.} \cite{Alves10} proposed a percentile-based method that computes thresholds by aggregating metric distributions across systems and selecting fixed percentiles to label risk levels from “low” to “very high”. This method involves weighting each metric value by the system size (in terms of LOC) and applying a multi-step aggregation process. However, this approach does not explicitly consider fault data and was mainly designed to enable benchmarking across projects.
Building upon this, Vale \textit{et al.} \cite{Vale19} proposed a benchmark-based methodology that improves risk granularity by associating thresholds with five risk levels. Their empirical validation showed better performance in detecting code smells compared to Alves' method \cite{Alves10}. Similarly, Ferreira \textit{et al.} \cite{FERREIRA2012} introduced a visual boxplot-based approach where thresholds are inferred by categorizing metric values into three frequency-based groups, i.e., good, regular, and bad, though the extraction criteria remain partially manual and underdefined. Another notable contribution is the Compliance Rate method by Oliveira et al. \cite{Oliveira14}, which defines relative thresholds by specifying the percentage of entities that should comply with a given metric limit. This technique incorporates penalty functions to select the optimal threshold values and emphasizes the notion of design rule conformance without directly labeling components.
From a statistical standpoint, Shatnawi introduced the use of univariate logistic regression to estimate thresholds that separate fault-prone and non-faulty modules. This approach identifies threshold values at predefined risk levels (VARL) and was among the first to adapt techniques from the medical domain to software engineering \cite{Shatnawi10}. In a subsequent evolution, Shatnawi also explored the use of Receiver Operating Characteristic (ROC) curves, selecting thresholds that maximize the sum of sensitivity and specificity \cite{Shatnawi10_2}. While ROC-based thresholds showed good discriminative power in severity classification, they performed less effectively in binary fault prediction settings. Expanding on these findings, Boucher and Badri \cite{BOUCHER18} systematically compared the predictive effectiveness of VARL, ROC-based cutoffs, and percentile-based rankings across several software systems. Their empirical results suggested that ROC-based thresholds generally outperformed both the other thresholding techniques and even some machine learning models in terms of predictive consistency.

Finally, a more recent domain-specific approach was proposed by Stojkovski et al.\cite{Stojkovski17}, who computed thresholds by categorizing Android applications into functional groups (e.g., games, multimedia) and then selecting either the mean or upper percentiles depending on the distribution normality. This method aims to contextualize thresholds according to the type of application being analyzed, improving interpretability and relevance for practitioners.

Taken together, these methods underscore the diversity of threshold derivation strategies in the literature, ranging from data-driven statistical models to visually guided techniques. However, as noted by Mori \textit{et al.} \cite{mori2018}, the lack of domain-independent thresholds limits their generalizability and applicability across industrial settings. This reinforces the need for context-aware and explainable methodologies, particularly those that can be adopted and validated in collaboration with practitioners.
\color{black}

\subsection{Strengths and Weaknesses of Statistical and Machine Learning Approaches in Fault Prediction}
\label{subsec:comparison}

To provide an overview of the key differences between statistical and machine learning approaches for software fault prediction, Table~\ref{tab:ml_vs_statistical} summarizes the main trade-offs across six dimensions: explainability, implementation cost, predictive performance, industrial applicability, ease of integration, and actionable feedback. This comparison draws upon insights reported in the literature and highlights the practical considerations that influence the adoption of each technique in real-world settings.

\begin{table}[H]
\centering
\footnotesize
\caption{Comparison between Statistical and Machine Learning Approaches for Fault Prediction}
\label{tab:ml_vs_statistical}
\begin{tabular}{|p{2.7cm}|p{3.5cm}|p{3.5cm}|p{1.75cm}|}
\hline
\textbf{Aspect} & \textbf{Statistical \hspace{0.9cm} Approaches} & \textbf{ML Approaches}  & \textbf{Reference} \\
\hline \hline
\textbf{Explainability} & \textbf{High}: models and thresholds are transparent and easy to understand & \textbf{Low}: models often behave as black boxes, hindering explainability & \cite{Haldar23},\cite{Santos2022},\cite{Begum2023}  \\
\hline
\textbf{Implementation Cost} & \textbf{Low}: requires minimal setup and no specialized infrastructure & \textbf{High}: requires infrastructure, training, and tuning & \cite{Stradowski24},\cite{STRADOWSKI23}\\
\hline
\textbf{Accuracy on Complex Data} & \textbf{Moderate}: struggles with non-linear patterns and large datasets & \textbf{High}: able to capture complex, non-linear relationships among data & \cite{BOUCHER18}, \cite{Pandey2021} \\
\hline
\textbf{Industrial \hspace{0.9cm} Applicability} & \textbf{High}: low setup effort and allows traceability and explainability needed in safety-critical domains & \textbf{Low}: lack of  industrial dataset, high setup effort and steep learning curve & \cite{rana14},\cite{Son19},\cite{BATOOL22}, \cite{STRADOWSKI23} \\
\hline
\textbf{Ease of \newline     Integration} & \textbf{High}: easy to plug into existing QA workflows & \textbf{Low}: QA workflow integration and model maintenance require expertise & \cite{PACHOULY22},\cite{rana14},\cite{Khatri22} \\
\hline
\textbf{Actionable Feedback} & \textbf{High}: developers can act directly on metrics exceeding thresholds & \textbf{Moderate}: lacks explicit guidance unless explainable AI techniques are used  & \cite{rana14},\cite{STRADOWSKI23},\cite{singh23}\\
\hline
\end{tabular}

\end{table}

The comparative overview offered by the table highlights why statistical methods, despite their lower predictive power in complex scenarios, remain attractive in industrial contexts that prioritize interpretability, cost-efficiency, and traceability.
\color{black}

Statistical techniques for fault prediction offer easy implementation, interpretability, and lower computational costs compared to machine learning (ML) and deep learning (DL) approaches. Statistical approach relies on well-established mathematical models, making them easier to understand and apply in industrial settings. However, they often assume data normality and homoscedasticity, which may not hold in real-world software systems. Furthermore, statistical techniques struggle to capture non-linear relationships and complex dependencies within software metrics, which can limit their predictive accuracy, particularly for large and intricate datasets \cite{Pandey2021}.

In contrast, ML/DL-based fault prediction models excel in handling large-scale, high-dimensional datasets and uncovering hidden patterns that traditional statistical methods might overlook. These models, particularly deep learning architectures, achieve superior accuracy by learning complex relationships among software metrics. However, their primary limitation is their lack of explainability, making it difficult to understand how predictions are generated. This black-box nature hinders adoption in industrial environments where developers require actionable insights to guide refactoring and defect management \cite{Pandey2021, Son19, Kalonia24, Rathore20171}. Consequently, ML/DL-based methods are rarely deployed in practice, with industries favoring interpretable statistical models such as those presented in \cite{Yu2002, Ostrand2005}.

To bridge this gap, researchers are investigating approaches that balance predictive accuracy and interpretability \cite{Mori2019}. However, while these methods improve understanding, they often lack concrete threshold values for software metrics, which are crucial for monitoring software quality and guiding refactoring efforts \cite{Sultan2021, Santos17, Alves10}. Additionally, most studies fail to account for contextual factors, treating thresholds as static rather than adapting them based on project characteristics \cite{Zhang13}. 

As noted by Mori \textit{et al.} \cite{mori2018}, the absence of domain-independent thresholds significantly reduces the practical applicability of software metrics.
Their findings indicate that generic threshold values often fail to reflect the unique structural characteristics and quality expectations of different software domains. Moreover, Mori \textit{et al}. argue that metric thresholds are typically derived from benchmark datasets that do not accurately represent the complexity or constraints of industrial systems. This results in thresholds that are either too strict or too lenient when applied in practice. The authors emphasize that thresholds must be tailored to project-specific characteristics to be actionable and trusted by practitioners. Their results further support the idea that domain-specific calibration is not only beneficial but necessary to improve the stability, relevance, and acceptance of metric-based quality evaluations in real-world development environments.

\color{black}

Although ML-based techniques have shown high predictive accuracy in academic studies, their industrial adoption remains limited due to several well-documented challenges. A central issue is the unfavorable cost-benefit ratio associated with their deployment in real-world settings. As highlighted by Stradowski \textit{et al.} \cite{Stradowski24}, implementing ML models often leads to considerable technical debt and requires substantial investments in infrastructure, training, integration, and long-term maintenance. These costs are rarely justified by the marginal improvements in prediction performance, especially in environments with tight development cycles and limited QA resources. Similarly, the work \cite{PACHOULY22, rana14} emphasizes that preparing an ML-based environment demands significant time and specialized knowledge, which many industrial teams cannot allocate. Moreover, Khatri \textit{et al.} \cite{Khatri22} notes that even with successful training, ML systems often suffer from high overhead and complexity during deployment, which can disrupt existing workflows.

Another limitation, underscored in \cite{Khatri22, Son19, BATOOL22}, concerns the scarcity of reliable industrial datasets. Most studies rely on publicly available datasets such as PROMISE, which do not reflect the complexity, structure, or constraints typical of proprietary industrial systems. This lack of real-world data restricts the generalizability and practical relevance of many ML-based models. Stradowski \textit{et. al} \cite{STRADOWSKI23}  further explains that industrial-grade data often suffers from inconsistencies, incomplete labeling, and lack of documentation, making them difficult to use for training robust ML models. As a result, there is a disconnect between what is achieved in research settings and what can be feasibly applied in practice.

The literature also identifies other practical barriers that hinder ML adoption in industry. These include challenges in evaluating models under realistic conditions, insufficient attention to class imbalance, poor explainability of predictions, limited documentation of deployment efforts, and a general lack of guidance on how to integrate ML predictions into established quality assurance workflows. Stradowski \textit{et. al} \cite{STRADOWSKI23} points out that most studies fail to address the need for transparency, traceability, and actionable feedback, factors that are critical in regulated domains such as automotive software development. Rana \textit{et al.} \cite{rana14} further stresses that organizational resistance and trust issues often prevent developers from relying on opaque predictive models, especially when predictions cannot be easily interpreted or validated through expert judgment. In contrast, threshold-based techniques not only offer clear decision boundaries but also provide actionable feedback. As noted by Sigh \textit{et al.} \cite{singh23}, these approaches allow developers to directly inspect and interpret metric values that exceed specific thresholds, enabling immediate and targeted improvements in the codebase. This characteristic supports faster feedback cycles and aligns well with continuous integration workflows.
Furthermore, Boucher \textit{et al.} \cite{BOUCHER18} conducted a comprehensive empirical study comparing threshold-based techniques with machine learning and clustering approaches for fault-proneness prediction. Their results show that threshold-based models not only achieved comparable performance to supervised learning techniques but also exhibited greater stability across different datasets. Notably, models based solely on thresholds often outperformed hybrid approaches that combined thresholding with ML or clustering, particularly in terms of consistency and simplicity. These findings suggest that threshold-based methods can serve as a robust and practical alternative for fault prediction, especially in industrial contexts where ease of integration and actionable feedback are critical.

Given these constraints, this study deliberately adopts a statistical threshold-based approach. Our method provides a lightweight, interpretable, and cost-effective alternative that does not require ML expertise or complex infrastructure. It produces results that are transparent and aligned with the traceability and documentation needs of ISO 26262-compliant workflows \cite{Amalfitano19}.

\color{black}

\section{The industrial context}
\label{sec:indContext}

The company we collaborated with develops embedded firmware for integrated memory systems in the automotive domain. In this safety-critical context, producing high-quality code that adheres to standards such as ISO 26262 is crucial \cite{Venkitachalam15, Santilli24}. 

The firmware, written in C, acts as a bridge between hardware and the operating system. The firmware is optimized for assuring real-time performance while operating within the constraints of limited hardware resources like restricted memory. Designed for Embedded MultiMedia Card (eMMC), the firmware provides memory access functionalities to external hosts.
\begin{figure}[h!]
	\centering
	\includegraphics[width=0.4\textwidth]{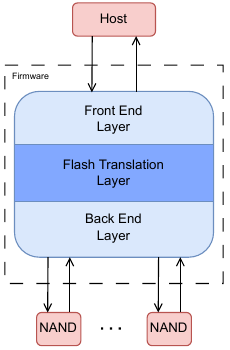}
	\caption{High-level Firmware Architecture}
	\label{fig:fw_arc}
\end{figure}

\subsection{Development Process and Artifact Traceability Model}
\label{sec:dev_process_traceability}

\begin{figure}[H]
	\centering
	\includegraphics[width=0.65\columnwidth]{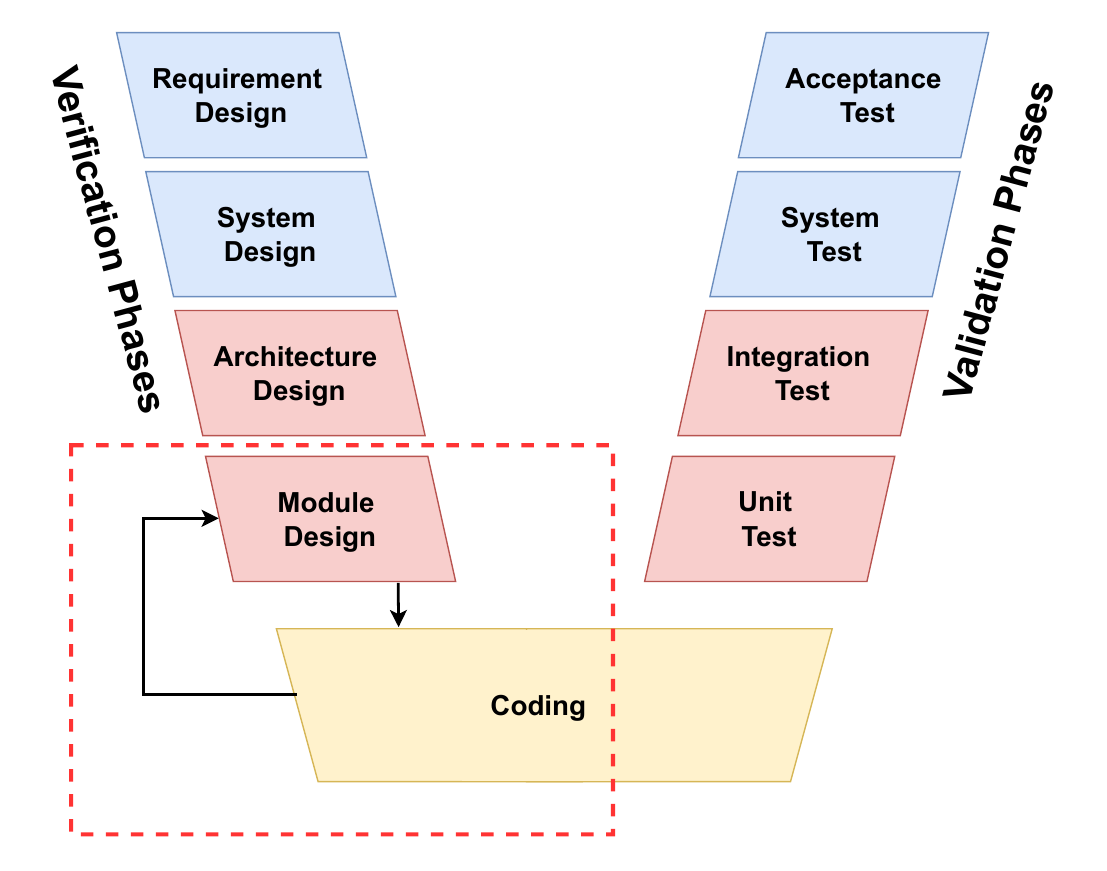}
	\caption{V-Model development process \cite{incose2015incose} with the highlighted area indicating where the proposed fault-prediction process is applied }
	\label{fig:dev_process}
\end{figure}

The company follows the V-Model as its software development process, as it represents the reference lifecycle model recommended by ISO 26262 for safety-critical systems \cite{iso26262-1-2018, Liu16}, as illustrated in Figure \ref{fig:dev_process}. While the development process is formally structured around the V-Model, the company also incorporates agile practices, such as iterative development and milestone-driven planning, to enhance flexibility and responsiveness. This hybrid approach supports frequent internal releases and facilitates early validation activities.

\begin{figure}[H]
\centering
\includegraphics[width=0.85\textwidth]{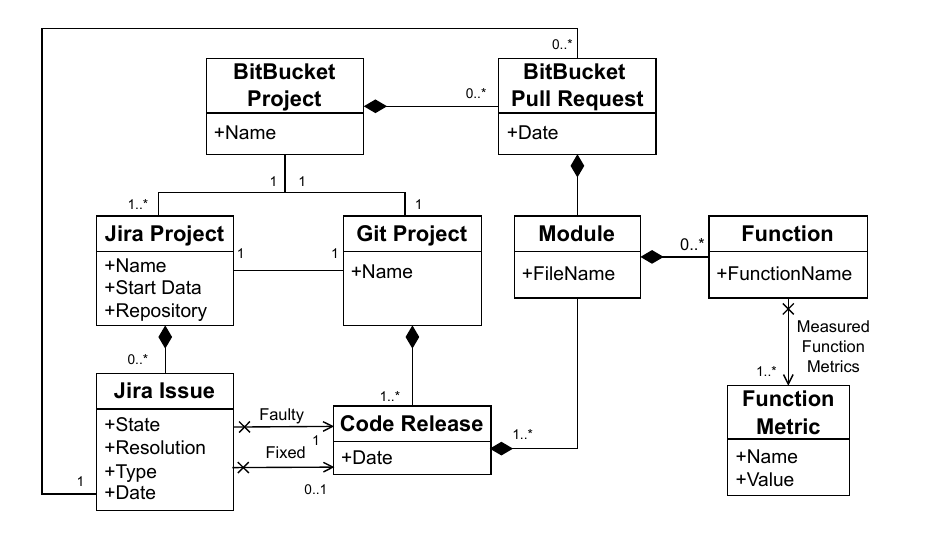}
\caption{Artifacts Traceability Model}
\label{fig:artefacts}
\end{figure}

Moreover, to reinforce software quality, the company conducts regular code review sessions where developers collaborate with senior engineers and project managers to evaluate critical modules. These reviews ensure that changes adhere to internal coding standards, maintainability goals, and safety-critical requirements. To support these SQA practices, the company employs two static analysis tools, \textit{Coverity Scan} and \textit{Understand}\footnote{\url{https://scan.coverity.com/}, \url{https://scitools.com/}}, which are integrated into the automated CI/CD pipeline. These tools provide a wide range of software metrics that are valuable for supporting compliance with ISO 26262 requirements.
\color{black}
Moreover, as required by the standard, the company ensures traceability between the various software artifacts throughout the development lifecycle \cite{Amalfitano19}. This includes maintaining bidirectional links between safety requirements, design models, source code, verification artifacts, and change requests. Such traceability facilitates both internal audits and external assessments, ensuring that each implemented feature or modification can be linked back to its originating requirement and associated safety rationale. This practice also supports efficient impact analysis, enabling teams to quickly determine which parts of the system must be re-verified when changes occur, thereby reducing the risk of unintended regressions. To better understand and formalize how development artifacts are related and how traceability is maintained, we conducted weekly one-hour focus groups with project managers and developers. These sessions allowed us to model the artifact relationships and gather insights into the tools and workflows adopted in practice. The UML class diagram in Figure~\ref{fig:artefacts} presents the structure of this traceability model.

The company employs \textit{Atlassian Jira} as project and task management software, which is used to manage backlogs, issues, bugs, and releases. It is used for issue tracking, particularly for faults identified during testing. \textit{Bitbucket} serves as the version control and pull request platform, integrated with \textit{Git} repositories.

In the model reported in Figure \ref{fig:artefacts}, a \textit{Bitbucket Project} manages one or more \textit{Jira Projects} and an associated \textit{Git Project}, and contains one or more \textit{Bitbucket Pull Requests}. Each pull request includes a release date and is linked to one or more \textit{Modules}. A \textit{Jira Project}, defined by its name, start date, and repository, is tied to a \textit{Git Project} and may contain multiple \textit{Jira Issues}. Each issue is described by its state, resolution, type, and date. Issues are linked both to the faulty code version and to the version where the fix was applied.
The \textit{Git Project} comprises one or more \textit{Code Releases}, each tagged with a release date and consisting of one or more \textit{Modules}. A \textit{Module} corresponds to a C source file and contains one or more \textit{Functions}. Each function is uniquely identified by its name and is associated with a set of measured software metrics. Each \textit{Function Metric} includes a name and a measured value, enabling fine-grained evaluation of software quality at the function level.

This detailed traceability framework plays a crucial role in linking quality metrics to actual fault data, enabling the empirical evaluation of metric thresholds and supporting continuous improvement in software quality assurance practices.
\color{black}
\subsection{Pain Points in the Software Quality Assurance Process}

In safety-critical domains, where development activities follow the V-Model, SQA activities such as requirement reviews, unit testing, integration testing, and system-level verification are distributed across design and implementation phases. However, as widely recognized in both academic and industrial literature, the later a defect is detected in this process, the higher its associated correction cost.

Delayed fault detection poses several practical challenges. Developers often have to address faults that emerge during system-level testing and originate from earlier design or coding decisions. Addressing these issues late in the cycle requires revisiting outdated code and risks invalidating previously completed documentation. 
Access to early indicators of potential defects in a module during the development process represents a viable solution to reduce the overhead associated with late defect discovery. 
Although static analysis tools such as \textit{Coverity} and \textit{Understand} are integrated into the company's CI/CD pipeline and produce a broad set of code metrics, these metrics are not currently exploited as indicators of potential risks or failures as they are not associated with threshold values. Without threshold values or contextual interpretation, developers struggle to understand whether a given function requires early refactoring, and quality issues may remain hidden until validation.

Although the ISO 26262 provides general principles for assessing the internal software quality, it does not specify measurement procedures or threshold values. Consequently, organizations must adapt quality attributes to their processes by selecting appropriate metrics, mapping them to quality goals (e.g., maintainability or complexity), and, eventually, defining threshold values to assess whether the software has achieved the required quality levels. Given the wide range of metrics across programming languages, development stages, and abstraction levels, this mapping is complex. Although many metrics are designed to be broadly applicable, they require contextual adaptation to be meaningful in specific industrial settings \cite{Vogel2021}.

To address this challenge, we decided to complement the company's SQA practices with a \textit{fault prediction technique based on metric thresholds}, which leverages the metric data already available in the company's workflow. This technique aims to identify fault-prone functions \textit{early in the development lifecycle}, particularly during the coding activities of the V-Model, where individual modules and functions are still under active development.
As highlighted in Figure~\ref{fig:dev_process}, the proposed fault prediction technique is designed to operate during module design and coding steps and before the validation activities, providing timely feedback to developers. This enables earlier and more focused quality assurance actions, supports fault prevention rather than fault detection, and aligns with the industry's need to strengthen early-phase verification.
In the following section, we present the proposed fault prediction process introduced to derive context-specific thresholds.

\color{black}

\section{Proposed Approach}
\label{sec:method_overview}

\begin{figure}[H]
    \centering
    \includegraphics[width=\textwidth]{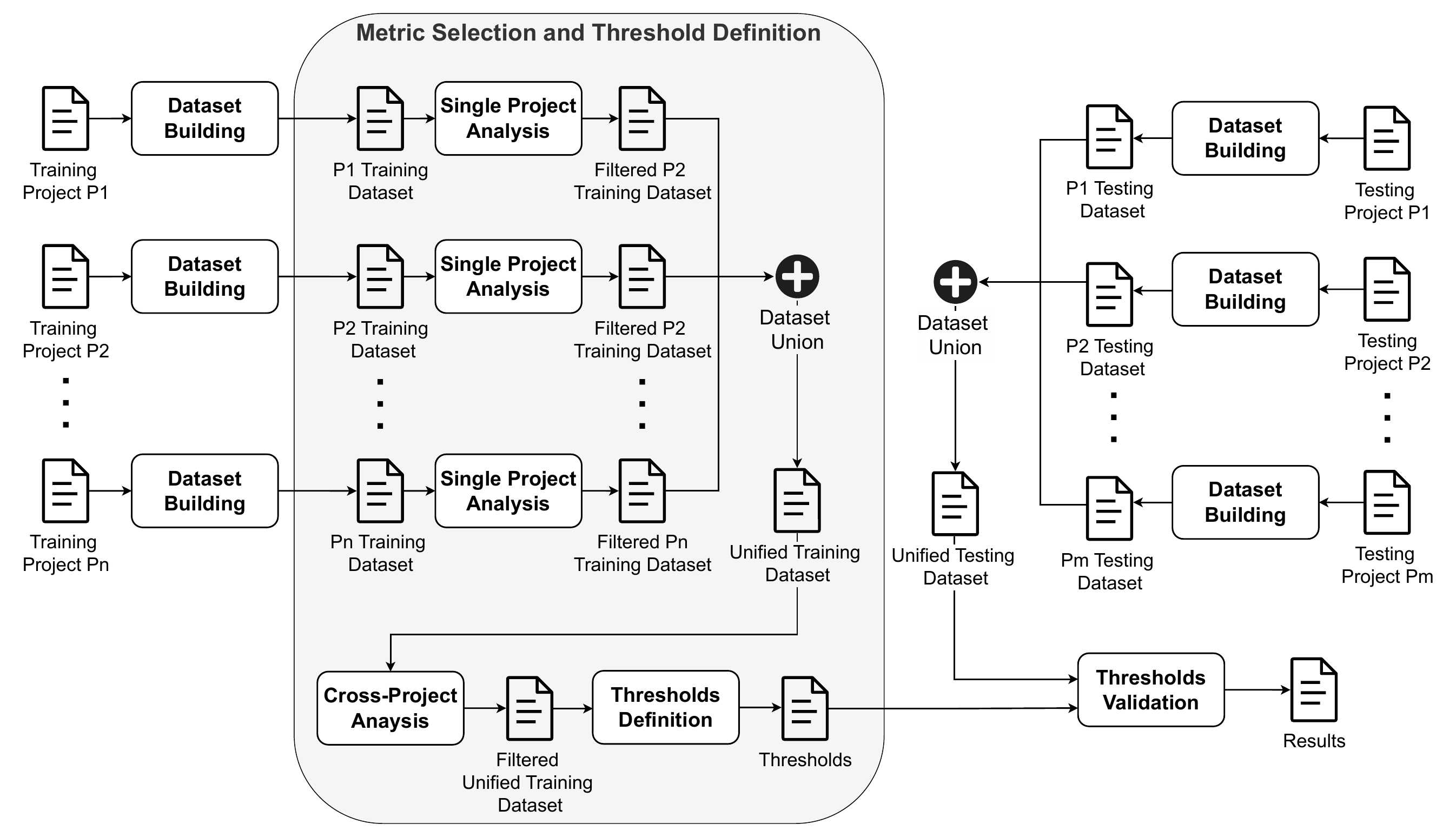}
    \caption{Process Overview}
    \label{fig:methodology_overview}
\end{figure}

The process, illustrated in Figure~\ref{fig:methodology_overview}, reflects the practical steps undertaken to define and validate metric-based thresholds for early fault detection in the considered industrial context.

Following established practices in cross-project defect prediction, the industrial software projects were divided into two distinct groups: training projects and testing projects \cite{Pal22}, a setup that mirrors practical industrial conditions where fault prediction models or thresholds are typically built from historical systems and applied to new projects during development.

The training projects (${P_1}$ to ${P_n}$) are used to derive thresholds, while the testing projects (also ${P_1}$ to ${P_m}$, but disjoint from the training set) are used to validate the effectiveness of these thresholds on unseen data.

The process begins with a \textit{Dataset Building} phase, in which metric datasets are constructed for both the training and testing projects. The datasets corresponding to the testing projects are immediately merged into a single \textit{Unified Testing Dataset}, which will be used later to validate the derived thresholds. In contrast, the datasets from the training projects are processed in advance and later merged in a \textit{Unified Training Dataset}, in the overall \textit{Metric Selection and Threshold Definition} phase.
This phase starts with a \textit{Single Project Analysis}, where each training dataset is individually analyzed to identify and remove metrics that either lack discriminative power or exhibit high correlation with each other. Once this filtering is completed for all training projects, the resulting datasets are merged. 
Next, a \textit{Cross-Project Analysis} is performed on the \textit{Unified Training Dataset} to identify and retain metrics that consistently differentiate faulty from non-faulty functions. Based on these identified metrics, the next step, \textit{Threshold Definition}, involves deriving empirical thresholds to flag potentially faulty functions. The final phase, \textit{Threshold Validation}, evaluates the generalizability and effectiveness of the derived thresholds by applying them to the unified testing dataset.

To implement this process, we utilized three real-world embedded firmware projects provided by our industrial partner. Two of these projects (${P_1}$ and ${P_2}$) were designated as training projects for threshold derivation, while the third (${P_3}$) served as an independent testing project.

The selection of these three projects was not arbitrary. We decided to include projects sharing similar structural characteristics, in terms of architectural design, codebase structure, and domain-specific functionalities, ensuring meaningful comparability.
This approach reflects a common strategy in the automotive domain and generally in safety-critical software development, where new firmware is rarely developed from scratch \cite{RUIZ17, Hardung04}. Instead, development typically builds upon validated components from previous systems to reduce risk and accelerate certification. By selecting structurally related projects, we ensure that the derived thresholds are both realistic and applicable to the early stages of development in similar future projects.

The following sections describe each of these five steps in detail and present the corresponding results.

\color{black}

\section{Dataset Building}
\label{sec:dataCollection}

In this section, we describe how the three datasets for the selected projects (${P_1}$, ${P_2}$, and ${P_3}$) were constructed. Specifically, the following sections present the Dataset Model (Section~\ref{subsec:DatasetModel}) and outline the process followed to build the dataset for each project (Section~\ref{subsec:DatasetBuilding}).

\color{black}
\subsection{Dataset Model}
\label{subsec:DatasetModel}

We define a dataset model to systematically represent software functions and their associated characteristics. This model provides a structured view of faulty and non-faulty functions, along with relevant software metrics, facilitating the analysis of fault-proneness at a function-grained level.

\begin{figure}[H]
	\centering
	\includegraphics[scale=0.9]{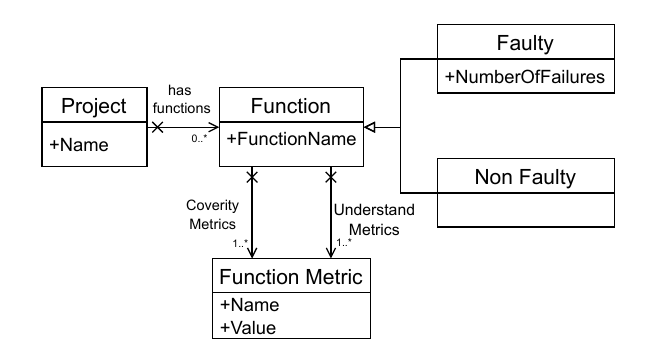}
	\caption{Dataset Model}
	\label{fig:dataset}
\end{figure}

The dataset model is illustrated in the UML class diagram shown in Figure \ref{fig:dataset}. As depicted, each \textit{Project} is uniquely identified by its name and consists of a collection of C functions. Each \textit{Function} is also identified by its name and is classified as either \textit{Faulty} or \textit{Non-Faulty}. For \textit{Faulty} functions, we document the number of times they have been reported as such through the attribute \texttt{NumberOfFailures}.
Although our analysis is based on a binary classification of functions into faulty and non-faulty, we included the \texttt{NumberOfFailures} attribute in the dataset model to support potential future analyses. This attribute, while not directly involved in the metric selection or threshold derivation process, enables further investigation into whether the recurrence of faults correlates with specific metric trends or whether it can be leveraged to model fault severity or fault density. Its inclusion preserves valuable historical fault information that could enrich future extensions of this study.
In addition to fault classification, each function is associated with two sets of measured metrics: one collected using Coverity and the other using Understand. These metrics, represented as key-value pairs of metric names and their corresponding values, provide detailed insights into the function's characteristics. By leveraging these metrics, we aim to capture critical aspects of code quality, supporting a more precise evaluation of fault-proneness at the function level. The function-level granularity of this model allows for fine-grained fault prediction, improving the identification of risk factors in software development.

\subsection{Dataset Building Process}
\label{subsec:DatasetBuilding}
\begin{figure}[H]
  \centering
  \includegraphics[width=1\textwidth]{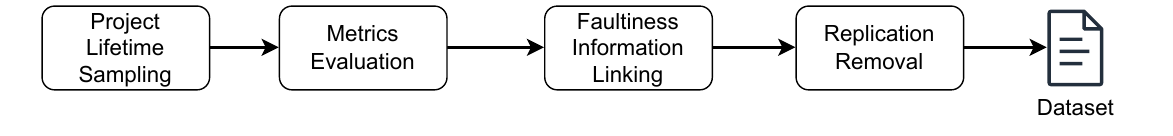}
  \caption{Dataset Building Process}  \label{fig:datasetbuild}
\end{figure}

The dataset construction process consists of four steps, as illustrated in Figure \ref{fig:datasetbuild}, and is designed to produce a dataset of faulty and non-faulty functions that adheres to the structure defined in the data model shown in Figure \ref{fig:dataset}. The following sections provide a detailed explanation of each step.

\subsubsection{Project Lifetime Sampling}

Discussions with two team leaders and a project manager from the development team highlighted that the codebase undergoes continuous modifications throughout the software development lifecycle. These changes stem from evolving stakeholder requirements and bug fixes identified during development. As a result, the codebase exhibits high variability, meaning that the version of the source code available at the end of development, marked by the Feature Development Complete (FDC) milestone, may differ significantly from earlier stages of the project.  The FDC milestone represents the point at which all planned features have been implemented, integrated, and tested.
It is used as a reference point to indicate that the system is functionally complete and ready to enter the final integration and validation phases of the development. Relying exclusively on functions available at the FDC commit would introduce two key issues:

\begin{enumerate}
    \item \textbf{Imbalance in the dataset}: Faulty functions, identified through \textit{Jira Issues}, can emerge and be resolved at any point during development. If we were to include only functions present at FDC, non-faulty functions would be constrained to a single point in time, whereas faulty functions would span the entire timeline. This discrepancy would result in an unbalanced dataset, where faulty functions are representative of the whole development process, while non-faulty functions are not. As a consequence, faulty functions would be overrepresented in the dataset, distorting the ability of the fault prediction process to generalize.
    \item \textbf{Lack of representativeness}: Limiting the dataset to functions available at FDC would fail to capture how functions evolve throughout the software lifecycle. Fault prediction is most effective when it accounts for changes occurring during development rather than being applied retrospectively. 
\end{enumerate}

\begin{figure}[H]
  \centering
  \includegraphics[width=0.9\textwidth]{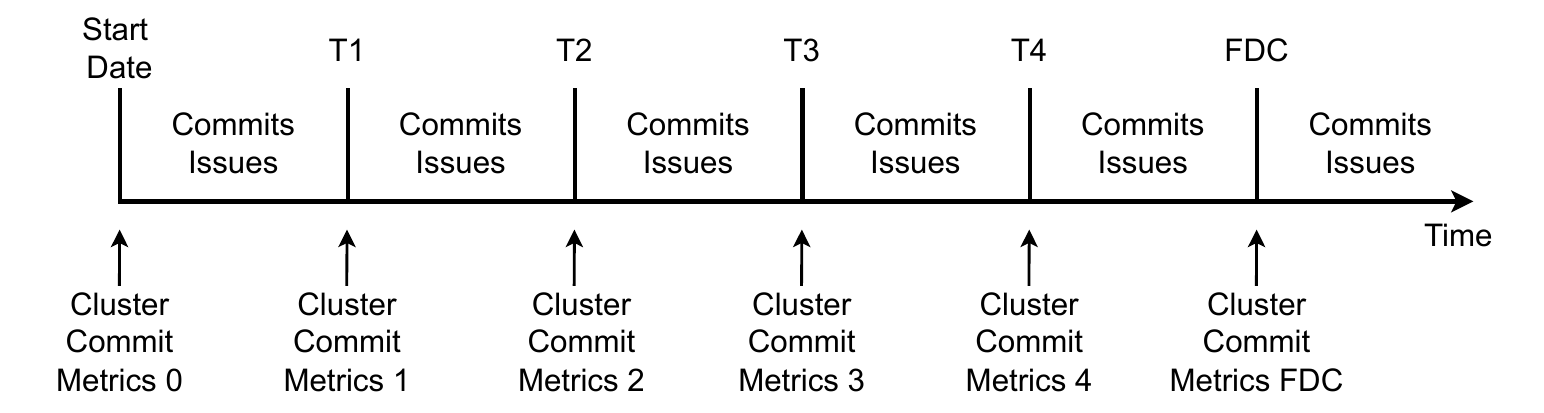}
  \caption{Project Lifetime Sampling}
  \label{fig:clustering}
\end{figure}

To address these challenges, we adopted a sampling strategy based on commit dates. 
\textcolor{black}{The project timeline was divided into $N=8$ intervals ($I_x$), each spanning two months as illustrated in Figure~\ref{fig:clustering}.} These intervals were selected to span representative phases of the project lifecycle, from the \textit{Start Date} to the \textit{FDC Date} following the internal release cadence. For each two-month interval, we selected the last commit preceding the interval boundary as the representative snapshot. This choice follows the company's internal practice of tagging stable builds at regular milestones and ensures that each snapshot corresponds to a validated, merge-complete state of the codebase.

This approach captures multiple snapshots of the codebase throughout development, mitigating the imbalance between faulty and non-faulty functions while ensuring that the dataset accurately reflects the software's evolution. Ultimately, this process resulted in 
\textcolor{black}{$N=8$ distinct snapshots of the codebase, each yielding a corresponding set of analyzed functions,}
where $N$ corresponds to the number of intervals used to sample the project lifetime.

\subsubsection{Metrics Evaluation}

In this step, we computed various metrics for the function sets collected in the previous phase using the Coverity and Understand tools. The set of metrics assessed by each tool is reported in Table \ref{tab:tools_metric}.

\begin{table}[t]
\centering
\caption{Coverity and Understand Metric}
\label{tab:tools_metric}
\begin{minipage}[t]{0.5\textwidth}
  \centering
  \includegraphics[width=\linewidth]{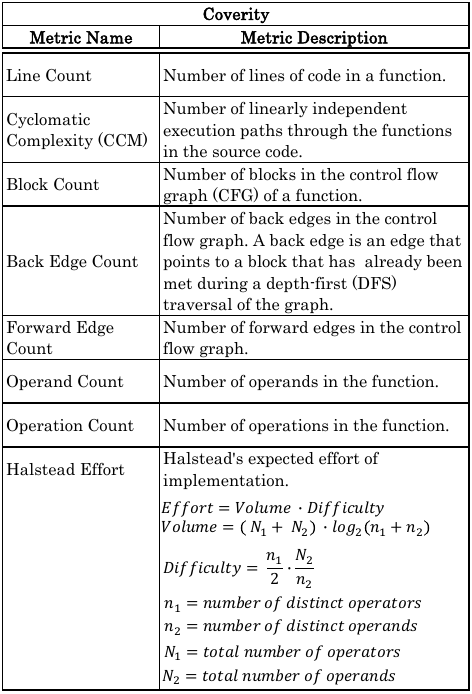}\\
  {\footnotesize (a) Coverity Metric}
\end{minipage}\hfill
\begin{minipage}[t]{0.49\textwidth}
  \centering
  \includegraphics[width=\linewidth]{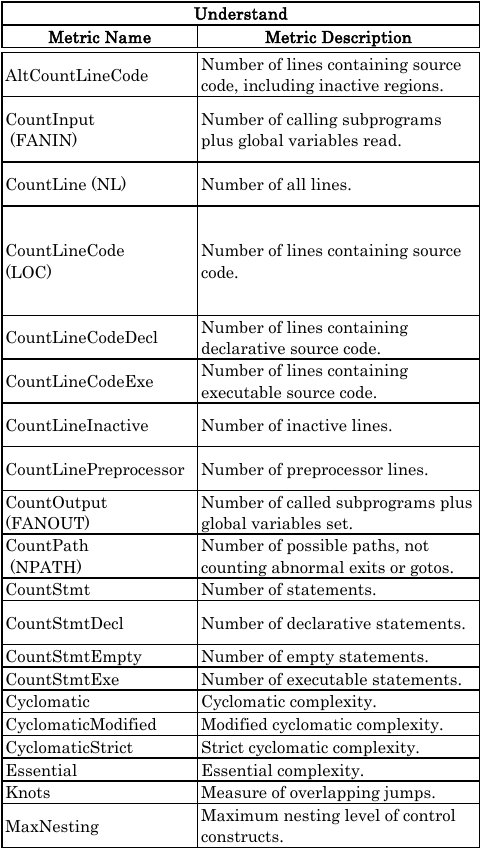}\\
  {\footnotesize (b) Understand Metric}
\end{minipage}
\end{table}

At the end of this step, the $N$ function sets were merged into a single dataset, with an example of this first intermediate dataset shown in Table \ref{tab:dataset_structure_intermediate_step1}. Each entry consists of the file path where the function is located, the function name, and the corresponding metric values extracted from both Coverity and Understand analyses. Additionally, a column is reserved for fault occurrences. However, at this stage, faulty information is not yet available, and it is therefore represented with a placeholder symbol ("-").

\begin{table}[H]
    \caption{Example of the first intermediate dataset}
    \label{tab:dataset_structure_intermediate_step1}
    \centering
    \includegraphics[width=0.85\textwidth]{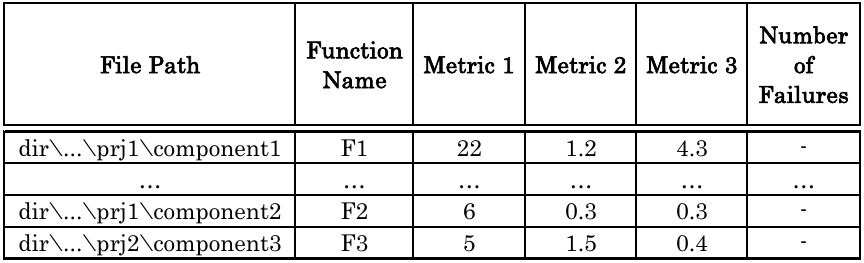}
\end{table}

\subsubsection{Faultiness Information Linking}

In this step, we identified faulty functions and recorded their fault occurrences. A function $F$ was classified as faulty based on the following assumption:  

\begin{tcolorbox}[colback=white!95!black, colframe=black, boxrule=0.5mm, arc=4mm]
\begin{quote}
    \textit{A function $F$ is considered faulty if it was modified to resolve a \textit{Jira Issue} opened during the software development process.}
\end{quote}
\end{tcolorbox}

The fault identification process followed these steps:  

\begin{enumerate}
    \item Using the artifacts traceability model (Figure~\ref{fig:artefacts}), we linked each \textit{Jira Issue} to its corresponding \textit{BitBucket Pull Request}.
    \item From the pull request, we identified functions modified to fix the issue.
    \item For each faulty function, we extracted its version from the commit immediately preceding the issue's opening date.
    \item Metrics were calculated on these extracted functions, ensuring they reflect the function's state before the bug fix.
\end{enumerate}

By applying this process across all \textit{Jira Issues}, we generated a comprehensive list of faulty functions with their associated metrics.  

At the end of this step, we obtained another intermediate dataset, building upon the previous representation. As shown in Table~\ref{tab:dataset_structure}, each entry includes the file path, function name, and calculated metrics. Unlike the earlier version, shown in Table \ref{tab:dataset_structure_intermediate_step1}, this dataset includes a value in the \texttt{NumberfFailures} field, set to 1 to indicate that the function was classified as faulty.

\begin{table}[h!]
        \caption{Example of the second intermediate dataset}
    \label{tab:dataset_structure}
    \centering
    \includegraphics[width=0.85\textwidth]{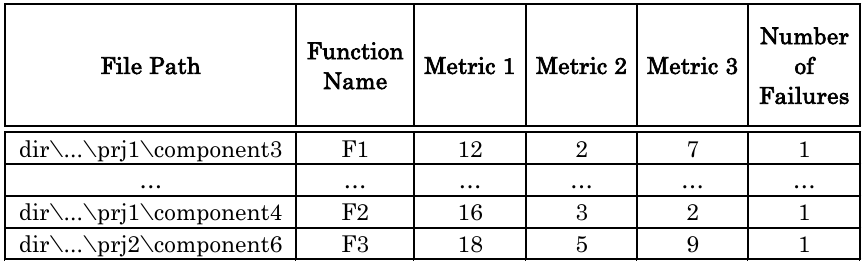}
\end{table}

\subsubsection{Removal of Replicated Metric Samples}

In this step, the two intermediate datasets, one representing non-faulty functions (Table~\ref{tab:dataset_structure_intermediate_step1}) and the other representing faulty functions (Table~\ref{tab:dataset_structure}), were merged into a single dataset. This merging process, along with the overall dataset construction, introduced duplicate entries that needed to be removed to ensure dataset integrity.  

The deduplication process followed these rules:  

\begin{enumerate}
    \item \textbf{Non-faulty functions:} If multiple samples had the same file path, function name, and identical metric values, only one was retained, and its \texttt{NumberOfFailures} attribute was set to $0$.  
    \item \textbf{Faulty functions:} If duplicate samples existed, only one was retained, and its \texttt{NumberOfFailures} attribute was updated to reflect the total fault occurrences across duplicates.  
    \item \textbf{Functions appearing as both faulty and non-faulty:} The faulty sample was retained, while the non-faulty sample was discarded.  
\end{enumerate}  

This process reduced dataset complexity while preserving essential information. 
Table~\ref{tab:dataset_structure_final} presents the final dataset structure, while 
\textcolor{black}{Table~\ref{tab:reducedDataset} reports the resulting number of retained samples per project after duplicate removal, and therefore represents the effective size of the dataset used in the subsequent analyses.}

\begin{table}[H]
    \caption{Example of the final dataset structure}
    \label{tab:dataset_structure_final}
    \centering
    \includegraphics[width=0.85\textwidth]{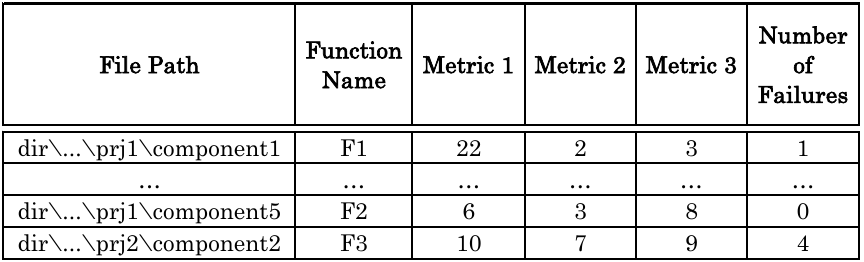}

\end{table}

\begin{table}[t]
\centering
\small
\caption{Number of samples after removing replicated data across the adopted tools}
\label{tab:reducedDataset}
\begin{tabular}{|l|c|c|}
\hline 
\textbf{Project} & \textbf{Coverity} & \textbf{Understand} \\
\hline \hline
Project 1 (${P_1}$) & 2,996 & 3,131 \\ \hline
Project 2 (${P_2}$) & 3,920 & 4,063 \\ \hline
Project 3 (${P_3}$) & 3,348 & 3,467 \\ \hline
\end{tabular}
\end{table}

\section{Metric Selection and Threshold Definition for Fault Prediction}
\label{sec:prediction}

This section presents the process used to evaluate the discriminative power of software metrics between faulty and non-faulty functions and the threshold extraction process to support cross-project fault prediction.
To guide the analysis, we inspired by the quantitative analysis process recommended by Wohlin \textit{et al.}~\cite{wohlin2024experimentation}. The process suggests a three-step approach; the first step involves performing a descriptive statistical analysis to explore the nature and distribution of the data. Insights gained during this step, together with those derived from correlation analysis, serve as input to the second step, which focuses on dataset reduction. This step aims to eliminate redundant or non-informative variables, streamlining the data for more effective interpretation. The filtered dataset is then used in the final step, where statistical hypothesis testing is applied to derive robust, data-driven conclusions.

\begin{figure}[h!]
    \centering
    \includegraphics[width=0.9\textwidth]{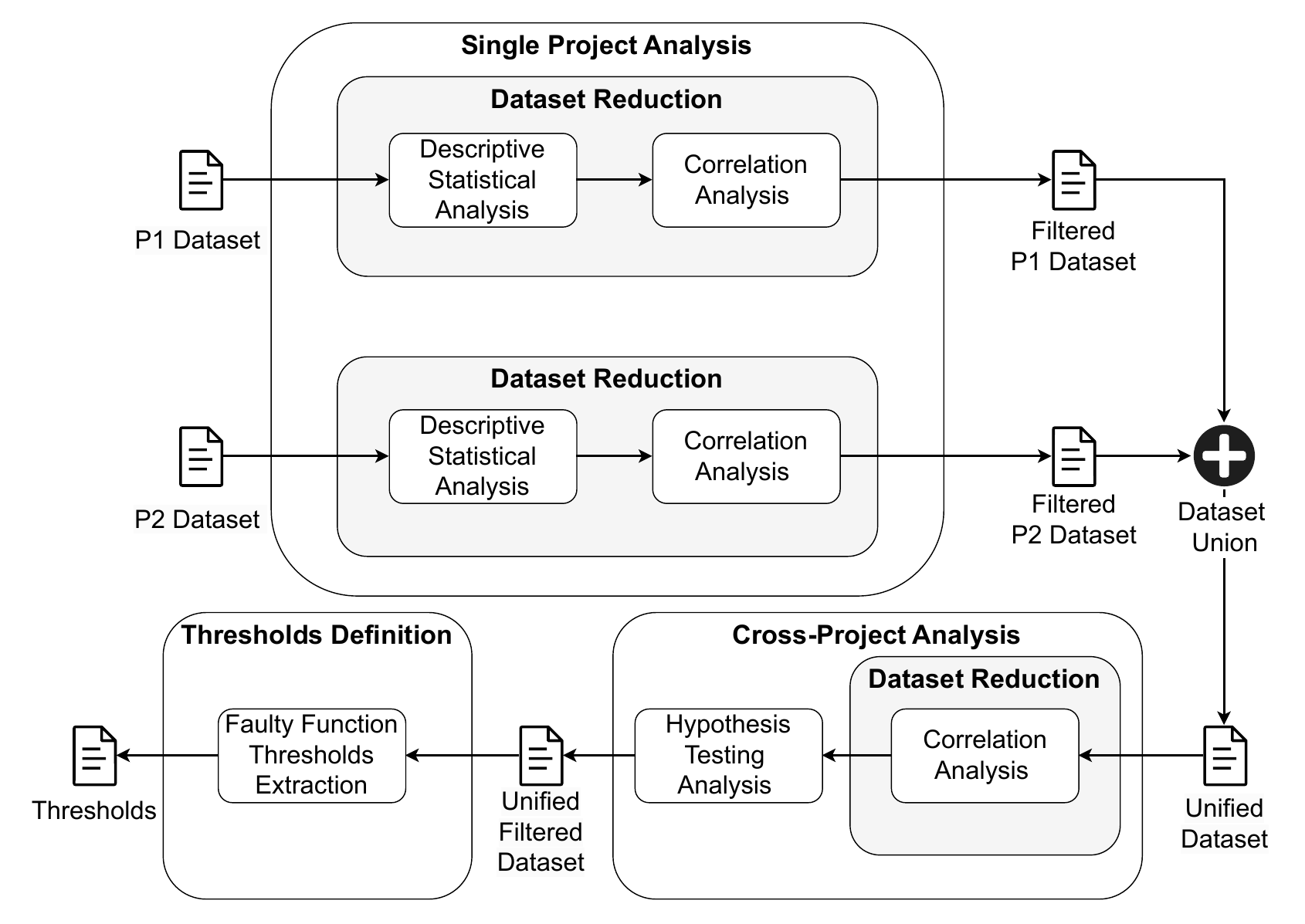}
    \caption{Proposed process for cross-project software fault prediction}
    \label{fig:pipeline}
\end{figure}

The proposed process, depicted in Figure~\ref{fig:pipeline}, is structured into three main steps, i.e., \textit{Single Project Analysis}, \textit{Cross-Project Analysis}, and \textit{Threshold Definition}.

The following sections detail the methodology employed in the analysis and present the key results obtained.

\subsection{Single Project Analysis}

In the \textit{Single Project Analysis}, each project's dataset was analyzed independently with the goal of performing \textit{dataset reduction}, by eliminating metrics that either lacked discriminative power between faulty and non-faulty functions or exhibited strong inter-correlation. The analysis consisted of two main steps:

\begin{itemize}
    \item \textit{Descriptive Statistical Analysis}: We conducted a visual analysis leveraging box plots and geometric density plots to visualize the distribution of each metric across the two classes. Metrics that did not show clear discriminative patterns, such as overlapping distributions or similar median values, were discarded.    
    \item \textit{Correlation Analysis}: Among the remaining metrics, we identified and removed those exhibiting high correlation. This step aimed to reduce redundancy and retain only independent and non-overlapping features that could contribute distinct information to the fault prediction model.
\end{itemize}

This two-step process allowed us to refine the metric set and retain only those features that were both discriminative and statistically independent, ensuring a robust foundation for the subsequent cross-project analysis and threshold definition.

\color{black}
In the following, we present the methodology using Project 1 (${P_1}$) as a representative example. The same approach was systematically applied to Project 2 (${P_2}$) for the Coverity and Understand static analysis tools.

\paragraph{Descriptive Statistical Analysis} 

\begin{figure}[h!]
\centering     
\subfigure[Box plot of \texttt{CCM} metric measured by Coverity tool of ${P_1}$]{\label{fig:CCM_box_plot}\includegraphics[width=60mm]{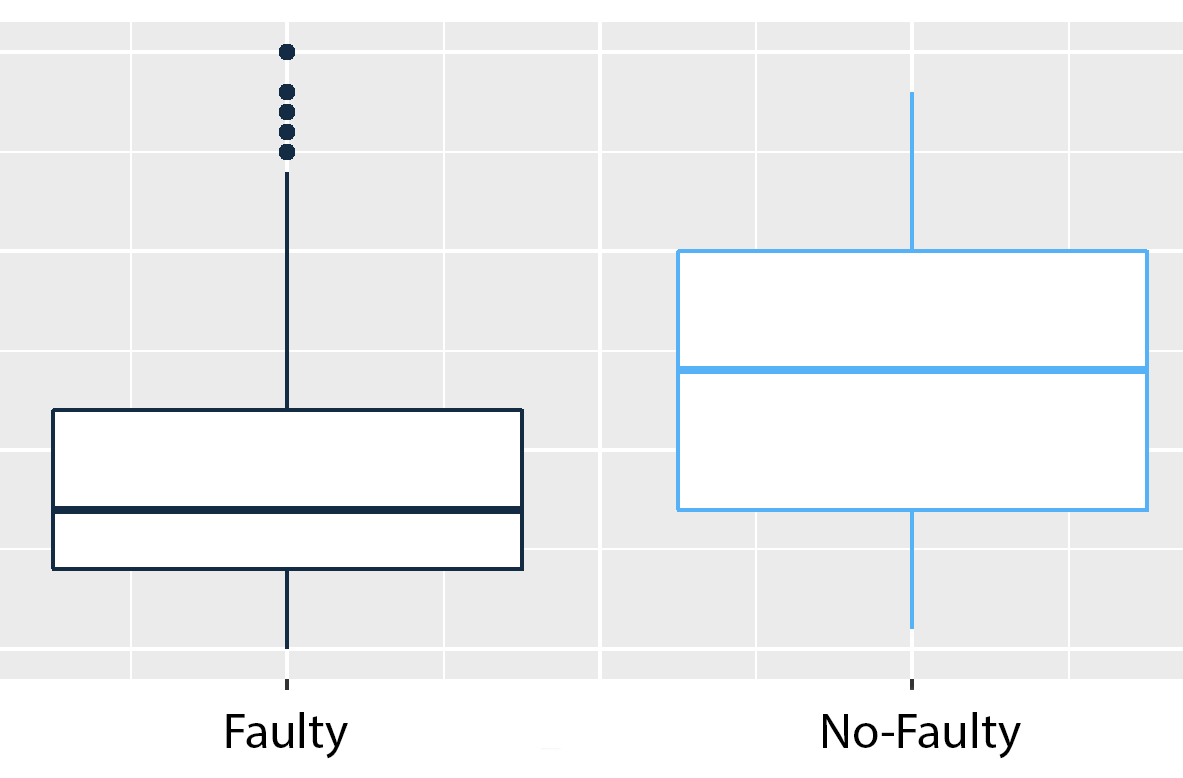}}
\subfigure[Geometric density of \texttt{CCM} metric measured by Coverity tool of ${P_1}$]{\label{fig:CCM_density}\includegraphics[width=60mm]{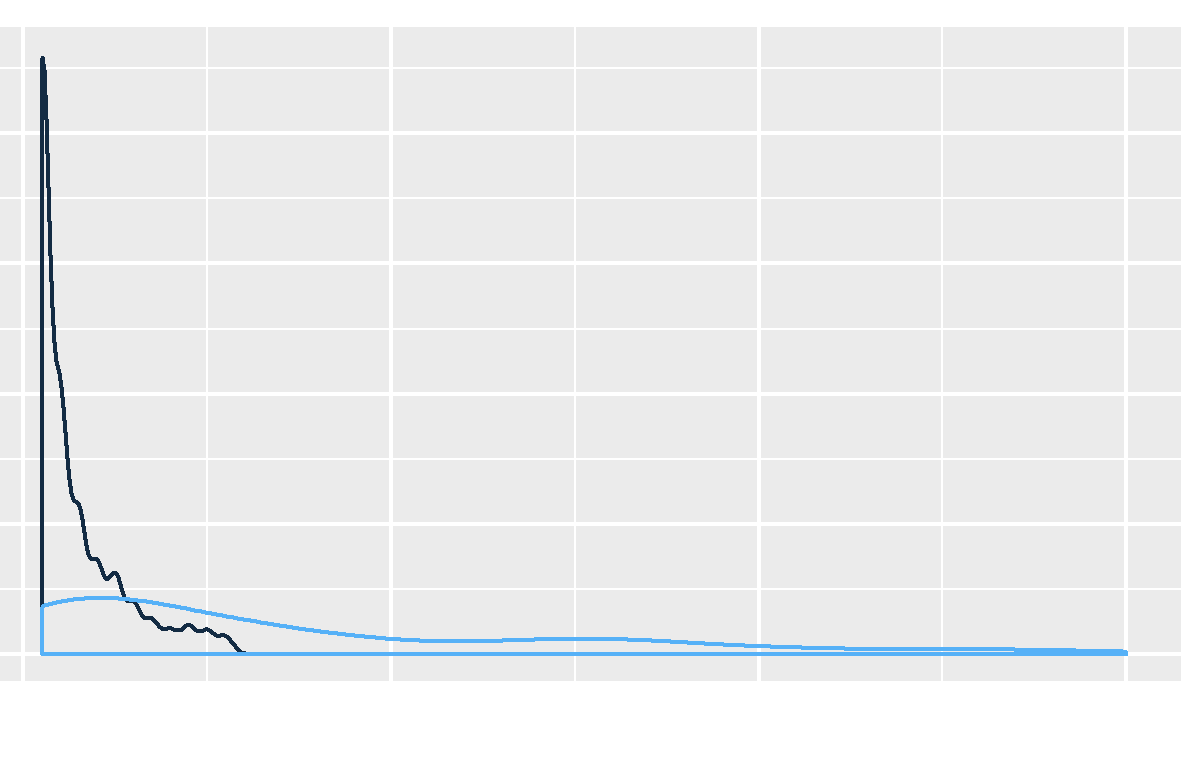}}
\subfigure[Box plot of \texttt{BackEdgeCount} metric measured by Coverity tool of ${P_1}$]{\label{fig:bec_box_plot}\includegraphics[width=60mm]{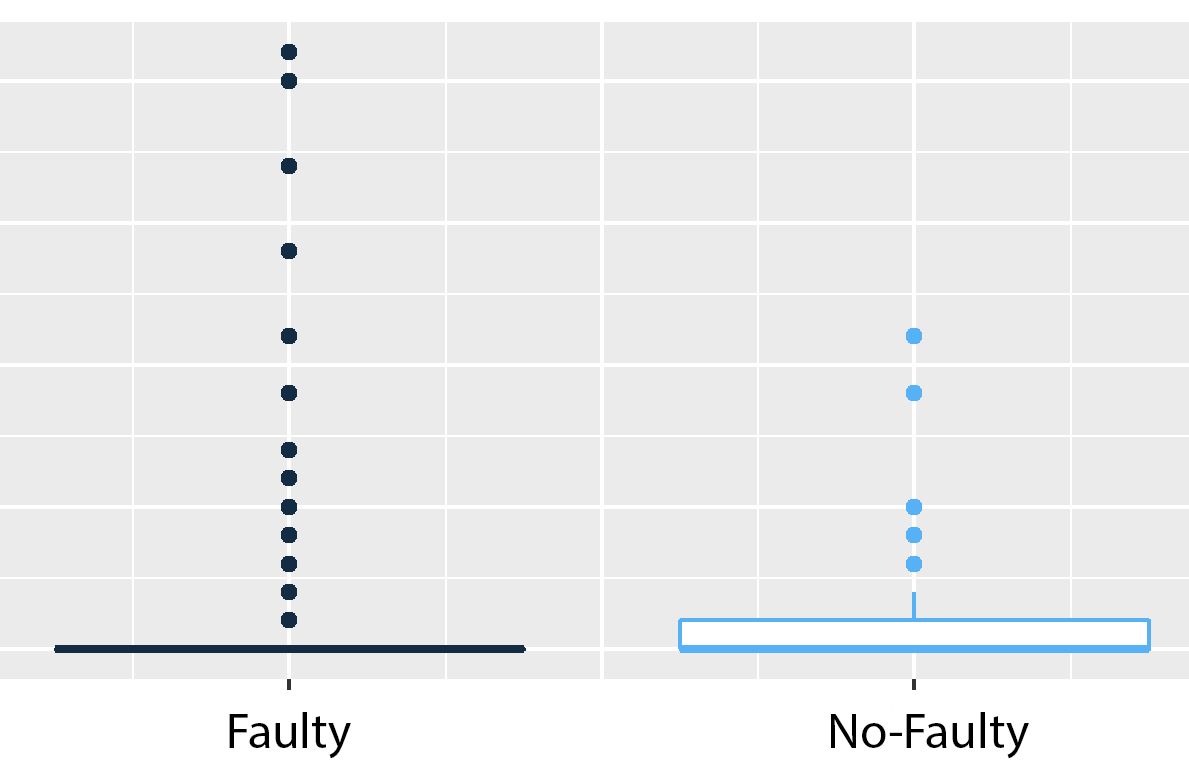}}
\subfigure[Geometric density of \texttt{BackEdgeCount} metric measured by Coverity tool of ${P_1}$]{\label{fig:bec_geo_density}\includegraphics[width=60mm]{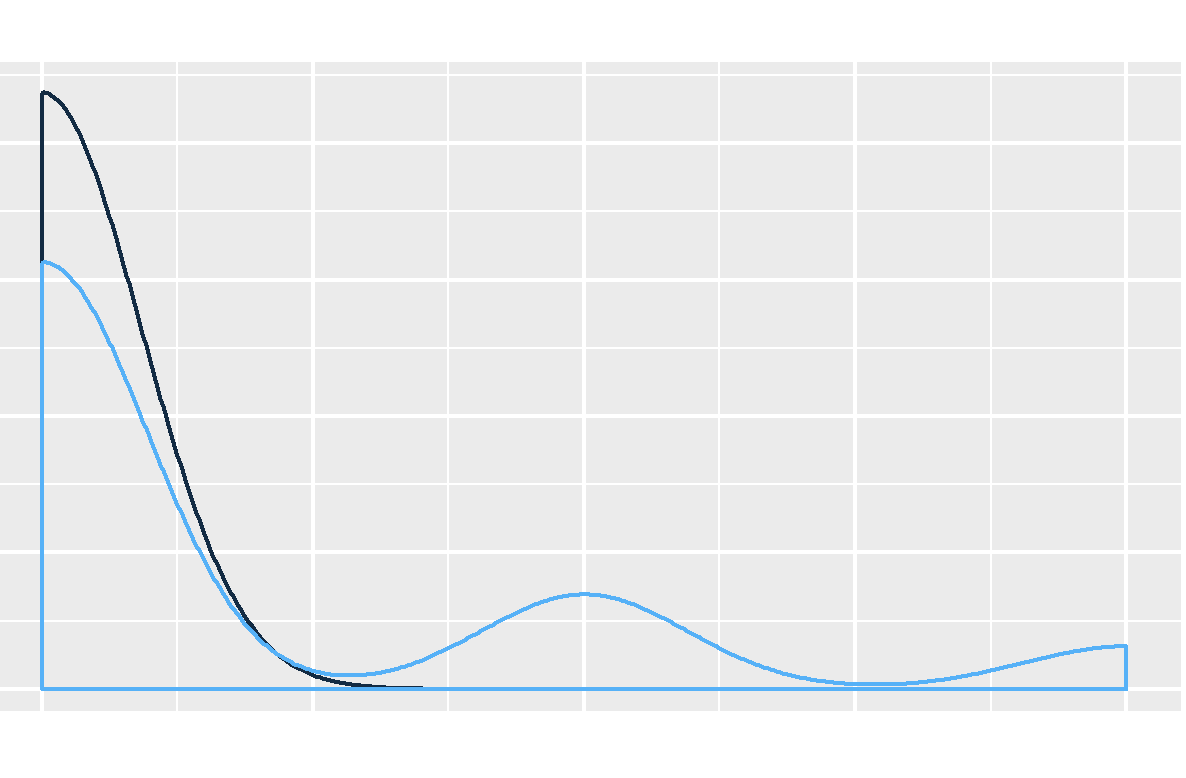}}
\caption{Faulty functions are represented in black ($\blacksquare$), while non-faulty functions are shown in blue (\textcolor{cyan}{$\blacksquare$}).}
\end{figure}

As a first step, we divided the ${P_1}$ dataset into two subsets: faulty functions and non-faulty functions. For each subset, we conducted a visual analysis using box plots and geometric density plots. These visualizations provide a more intuitive understanding of how metric distributions differ between faulty and non-faulty functions. Figures \ref{fig:CCM_box_plot} and \ref{fig:CCM_density} illustrate the distribution of the \texttt{CCM} metric. Due to confidentiality agreements, axis values are omitted from the figures, as the underlying numerical information could potentially reveal sensitive details about the quality of the analyzed code. The box plot highlights a clear difference in median values between faulty and non-faulty functions, while the density plot shows minimal overlap between their distributions, suggesting a strong discriminative power. Based on this observation, we retained \texttt{CCM} in our dataset. In contrast, Figures \ref{fig:bec_box_plot} and \ref{fig:bec_geo_density} depict the distribution of \texttt{BackEdgeCount}, which exhibits significant overlap and no notable difference in median values. This indicates its limited effectiveness in distinguishing between the two categories, supporting its exclusion from the dataset.

\paragraph{Correlation Analysis and Metric Selection}
In this second step, we computed Pearson's correlation on the filtered set of metrics obtained from the previous step to retain only independent and non-redundant metrics. 
To identify and remove highly correlated metrics, we computed Pearson's correlation coefficient and applied a threshold of 0.9. 
While correlation values above 0.5 are often considered indicative of moderate to strong linear relationships, we adopted a conservative threshold of 0.9 to filter only clearly redundant metrics. 
This choice was motivated by the need to preserve a sufficiently diverse and interpretable set of metrics for practical use in industrial settings. In our context, metrics with a correlation between 0.5 and 0.9 were not assumed to be independent, but were instead retained for further evaluation through structured focus group sessions with senior developers and project managers. These practitioners selected the most intuitive and actionable metrics among the correlated ones, ensuring that the final selection was both statistically justified and aligned with the software quality assurance needs of the company. 
\textcolor{black}{
Overall, Pearson correlation with a high cutoff was deliberately employed as a conservative pruning mechanism aimed at identifying only obvious redundancy, rather than modeling fine-grained dependency structures among metrics.
Given the discrete and skewed nature of the considered metrics, this design choice was complemented by a rank-based sensitivity check (e.g., Spearman and Kendall) on the unified training dataset.}
\color{black}
Table \ref{tab:prj1_CorrelationAll} presents the Pearson correlation coefficient among the selected Coverity metrics for ${P_1}$. We observed strong correlations among \texttt{LineCount}, \texttt{BlockCount}, \texttt{CCM}, and \texttt{ForwardEdgeCount}, as well as between \texttt{HalsteadEffort} and \texttt{BlockCount}. To determine which metrics to retain among those found to be highly correlated, we conducted focus group sessions involving senior developers and project managers from the company. In these sessions, the practitioners themselves selected the metrics they considered more intuitive, interpretable, and easier to monitor and act upon in their quality assurance tasks. In our case, \texttt{ForwardEdgeCount}, \texttt{BlockCount}, and \texttt{LineCount} were removed, and \texttt{CCM} was retained.

\begin{table}[h!]
    \caption{Project 1 (${P_1}$) Coverity Metrics Pearson Correlation Matrix}
    \label{tab:prj1_CorrelationAll}
    \centering
    \includegraphics[width=0.9\textwidth]{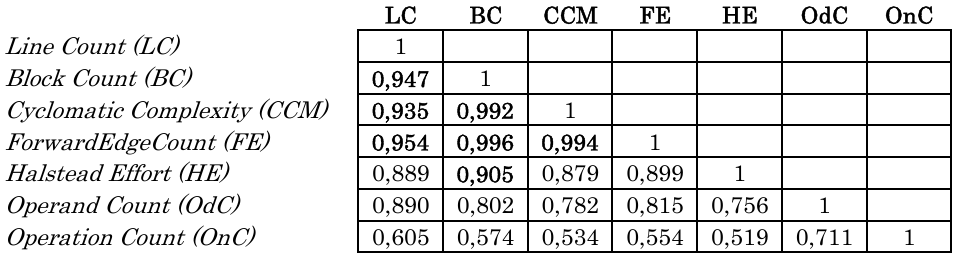}

\end{table}

\paragraph{Extension to Understand Metrics}

\begin{figure}[h]
\centering     
\subfigure[Box plot of \texttt{CountInput} metric measured by Understand tool of ${P_1}$]{\label{fig:understand_box_plot_1}\includegraphics[width=60mm]{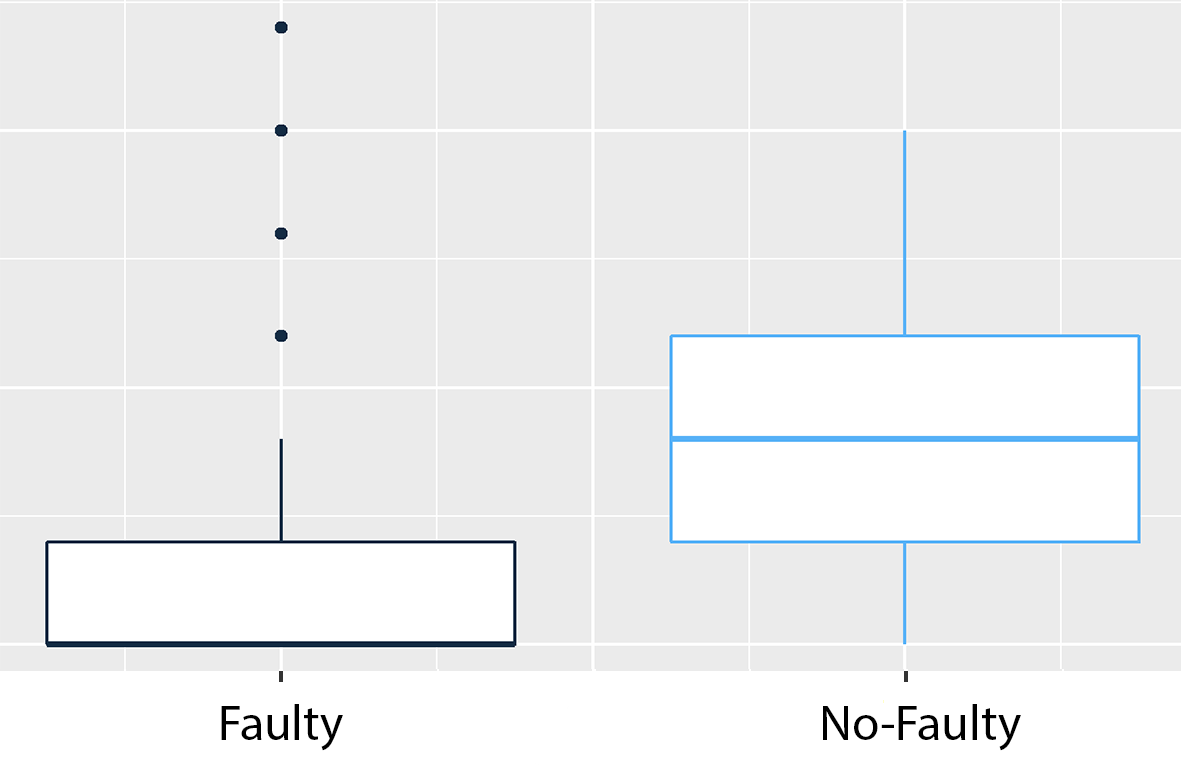}}
\subfigure[Geometric density of \texttt{CountInput} metric measured by Understand tool of ${P_1}$]{\label{fig:undestand_density1}\includegraphics[width=63mm]{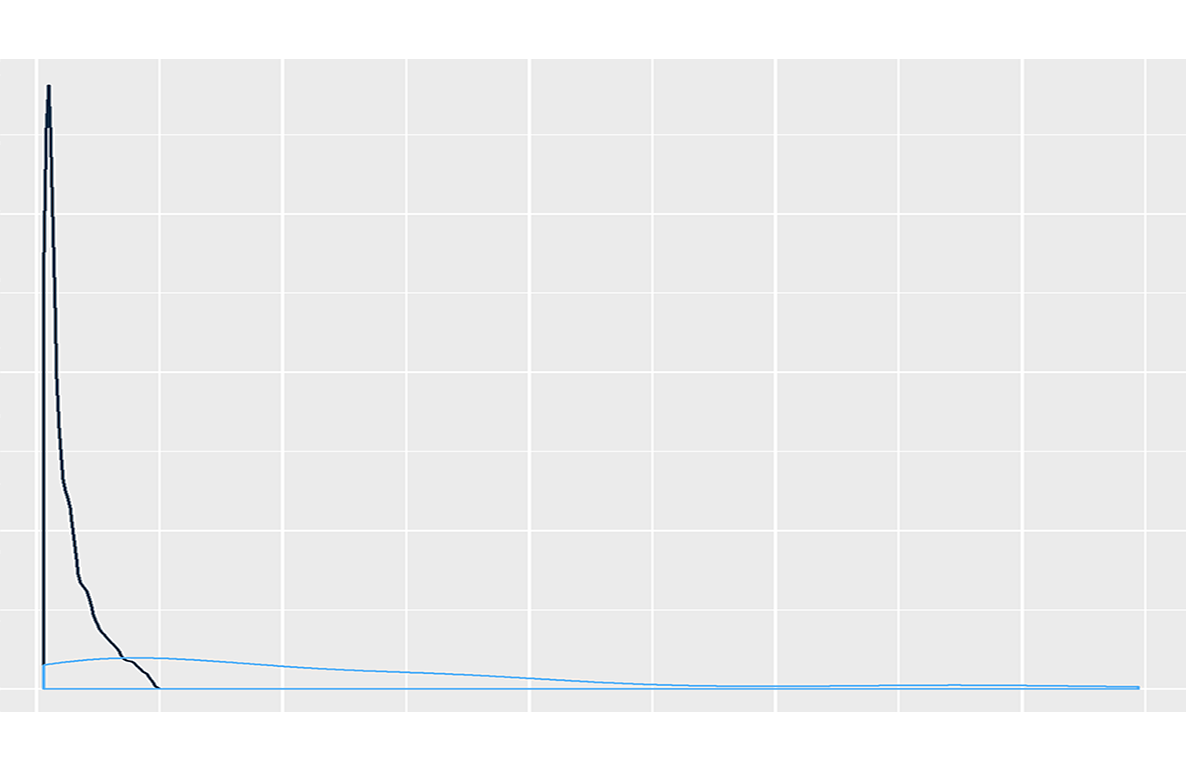}}
\subfigure[Box plot of \texttt{CountPath} metric measured by Understand tool of ${P_1}$]{\label{fig:bec_box_plot_1}\includegraphics[width=60mm]{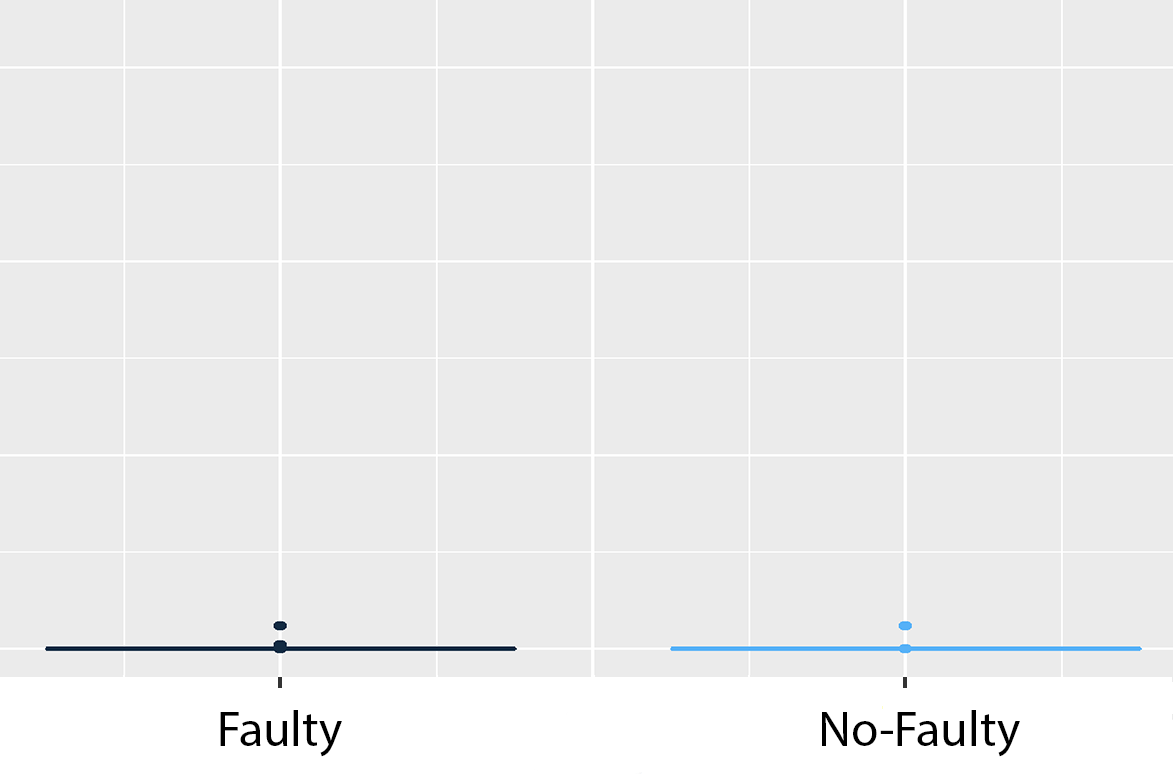}}
\subfigure[Geometric density of \texttt{CountPath} metric measured by Understand tool of ${P_1}$]{\label{fig:bec_geo_density_1}\includegraphics[width=63mm]{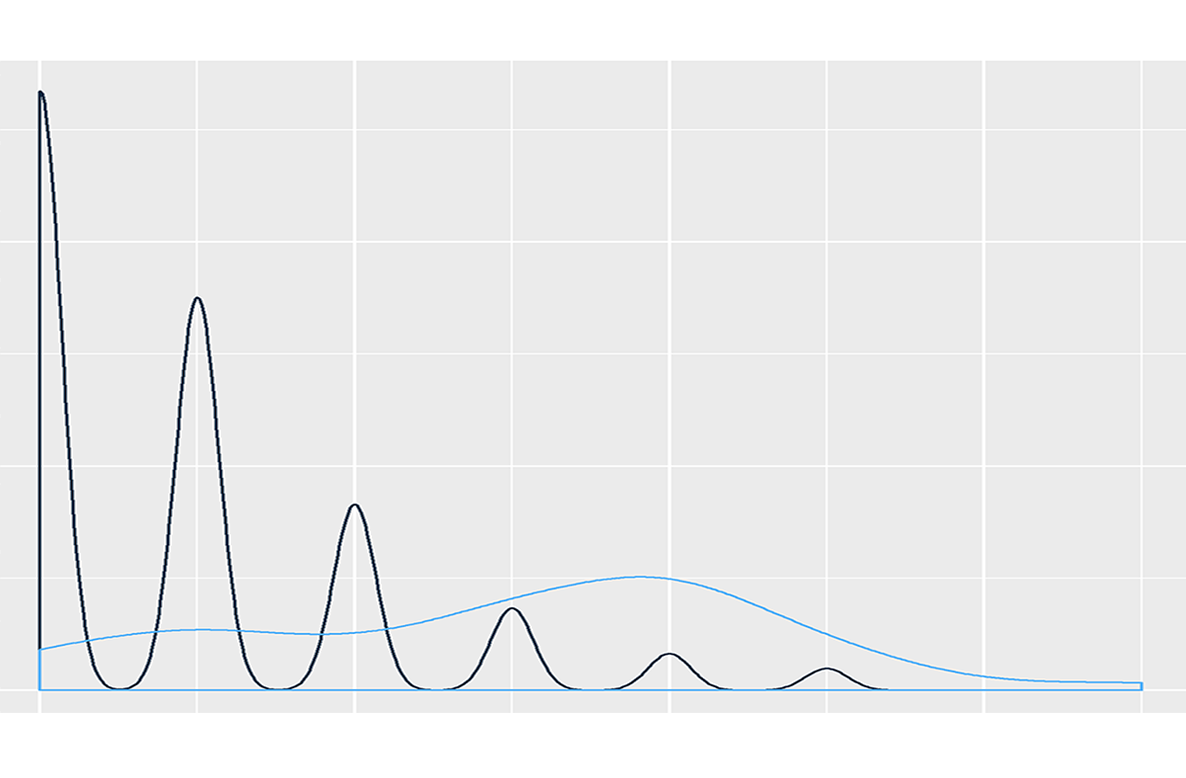}}
\caption{Faulty functions are represented in black ($\blacksquare$), while non-faulty functions are shown in blue (\textcolor{cyan}{$\blacksquare$}).}
\label{fig:understand_visual}
\end{figure}

The same methodology was systematically applied to the metrics extracted using the Understand static analysis tool. As with Coverity, the evaluation began with a visual inspection to assess the discriminatory power of each metric in separating faulty from non-faulty functions. Metrics such as \texttt{CountLinePreprocessor}, \texttt{CountLineInactive}, \texttt{CountPath}, \texttt{CountStmtEmpty}, and \texttt{CountStmtDecl} exhibited substantial distributional overlap between faulty and non-faulty functions, with no notable differences in median values. As illustrated in Figure \ref{fig:understand_visual}, these metrics failed to provide useful separation and were consequently excluded from further analysis due to their limited discriminative capability. Following the visual inspection, Pearson's correlation analysis was performed to identify and remove redundant metrics. A threshold of 0.9 was again applied to identify pairs of highly correlated metrics.

As reported in Table \ref{tab:prj1_CorrelationAll_understand}, several metrics showed strong correlation with \texttt{CountLineCodeExe}, including \texttt{AltCountLine}, \texttt{CountLine}, \texttt{CountLineCode}, \texttt{CountStatement}, and \texttt{CountStatementExe}. Additionally, \texttt{CountLineCodeDecl} and \texttt{CountLineCode} were found to be strongly correlated, resulting in the exclusion of the latter. To finalize the selection among correlated candidates, structured focus group sessions were again held with senior developers and project managers. The participants selected the metrics they found most intuitive and actionable within their software quality workflows. Their input was used to guide the retention of metrics such as \texttt{CountLineCodeExe} and \texttt{CountLineCodeDecl}, which were perceived as both meaningful and practically useful in real-world quality assessment activities.
\color{black}

\begin{table}[h!]
    \caption{Project 1 (${P_1}$) Understand Metrics Pearson Correlation Matrix}
    \label{tab:prj1_CorrelationAll_understand}
    \centering
    \includegraphics[width=\textwidth]{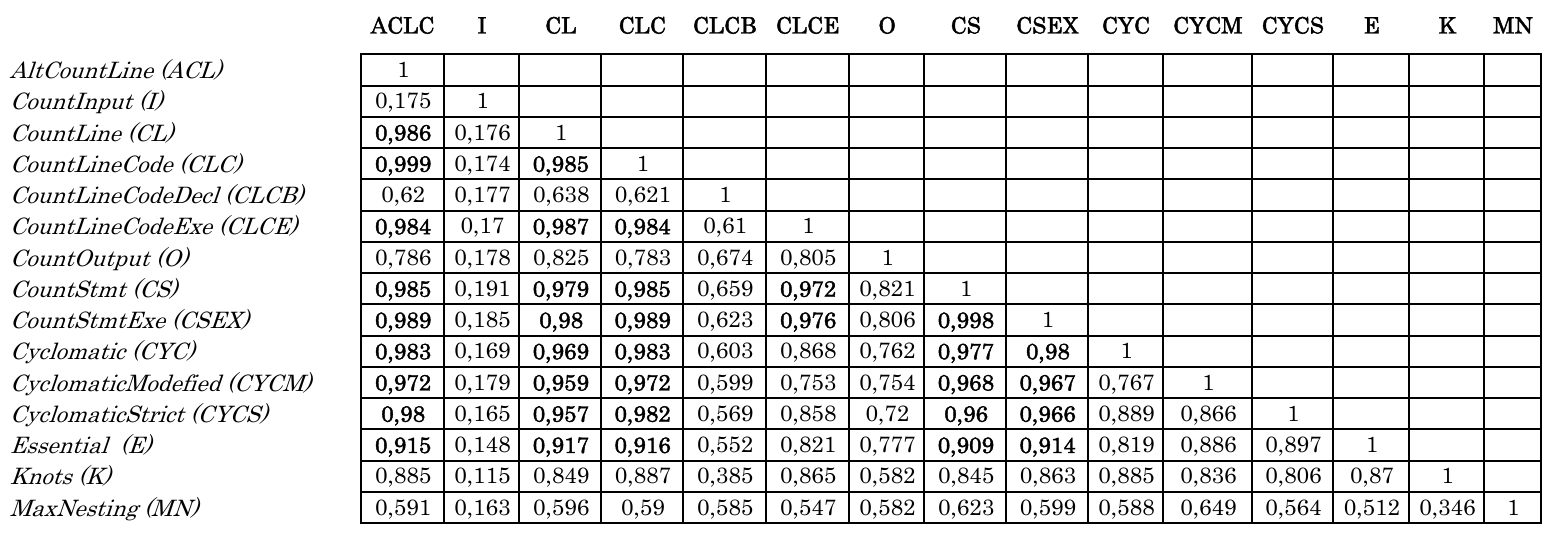}

\end{table}

\paragraph{Summary of Metric Selection Across Projects}

Through visual inspection and correlation filtering, we refined the set of discriminative metrics across the projects, removing those that failed to clearly distinguish between faulty and non-faulty functions or exhibited high collinearity. This process resulted in the same set of filtered metrics for all the projects, presented in Table \ref{tab:selected_metrics}. At the end of this step, we obtained the Filtered Dataset for each project, as reported in Figure \ref{fig:pipeline}.

\begin{table}[t]
\centering
\caption{Selected Coverity and Understand Metrics}
\label{tab:selected_metrics}

\begin{minipage}[t]{0.42\textwidth}
\centering
\small
\caption*{(a) Selected Coverity Metrics}
\begin{tabular}{|l|}
\hline
\textbf{Coverity Metric} \\
\hline \hline
CCM \\ \hline
Halstead Effort \\ \hline
Operand Count \\ \hline
Operation Count \\ 
\hline
\end{tabular}
\end{minipage}
\hfill
\begin{minipage}[t]{0.48\textwidth}
\centering
\small
\caption*{(b) Selected Understand Metrics}
\begin{tabular}{|l|}
\hline 
\textbf{Understand Metric} \\
\hline \hline
CountInput \\ \hline
CountLineCodeDecl \\ \hline
CountLineCodeExe \\ \hline
CountOutput \\ \hline 
Cyclomatic \\ \hline
CyclomaticModified \\ \hline
CyclomaticStrict \\ \hline
Essential \\ \hline
Knots \\ \hline
MaxNesting \\ \hline
\end{tabular}
\end{minipage}

\end{table}

\subsection{Cross-Project Analysis}

In the \textit{Cross-Project Analysis}, we combined the datasets from the two training projects, ${P_1}$ and ${P_2}$, into a unified dataset to evaluate the generalizability of the selected metrics in a cross-project setting. This analysis was structured into two step. First, we verified that merging the datasets did not introduce strong correlations among the metrics, thereby preserving their independence. We then applied statistical hypothesis testing, conducted separately for the Coverity and Understand metrics, to assess each metric's ability to distinguish between faulty and non-faulty functions across the merged dataset. This separation allowed us to account for the specific characteristics and distributions of each tool's metric set.

In the following, we detail each step of the cross-project analysis.

\subsubsection{Unified Dataset Correlation Analysis}
\color{black}

\textcolor{black}{
In this step, to verify potential redundancies among the selected metrics after merging the datasets, we analyzed pairwise associations on the unified training dataset by computing correlation coefficients for all metrics extracted from Coverity and Understand separately.
As a conservative secondary pruning step aimed at identifying only obvious redundancy, Pearson's correlation coefficient was first applied with a high cutoff value ($r = 0.9$).
In addition to Pearson's correlation, we also computed Spearman's $\rho$ and Kendall's $\tau$ to perform a rank-based sensitivity analysis, as these measures are more appropriate for discrete and skewed metric distributions.
}

For Coverity, the correlation matrix in Table~\ref{tab:CoverityCorrelationMatrix} confirms the absence of correlations among the selected metrics, aligning with the findings from the individual dataset analysis.
\textcolor{black}{
Consistently, neither Spearman's nor Kendall's correlation analyses identified any metric pairs exceeding the adopted cutoff, confirming that no additional redundancy emerges under rank-based association measures for Coverity metrics.
}

\begin{table}[h!]
    \caption{Pearson Correlation Matrix for Coverity Metrics in the Unified Dataset}
    \label{tab:CoverityCorrelationMatrix}
    \centering
    \includegraphics[scale=0.8]{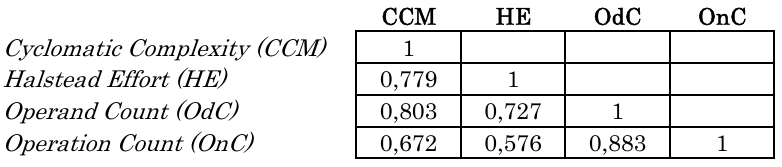}

\end{table}

Similarly, for Understand, the previously filtered metrics remain uncorrelated when considering Pearson's correlation after merging the datasets, as shown in Table~\ref{tab:UnderstandCorrelationMatrix}.
\textcolor{black}{
However, the rank-based analyses revealed additional high correlations among a subset of Understand metrics.
Table~\ref{tab:UnderstandSpearmanCorrelation} reports the metric pairs whose Spearman correlation exceeds the cutoff, highlighting associations among cyclomatic metrics (\texttt{Cyclomatic}, \texttt{CyclomaticStrict}, and \texttt{CyclomaticModified}), and their relationship with \texttt{MaxNesting}. A weaker, yet above-threshold, association was also observed between \texttt{Cyclomatic} and \texttt{Knots}.
}

\begin{table}[H]
    \caption{Pearson Correlation Matrix for Understand Metrics in the Unified Dataset}
    \label{tab:UnderstandCorrelationMatrix}
    \centering
    \includegraphics[width=0.97\textwidth]{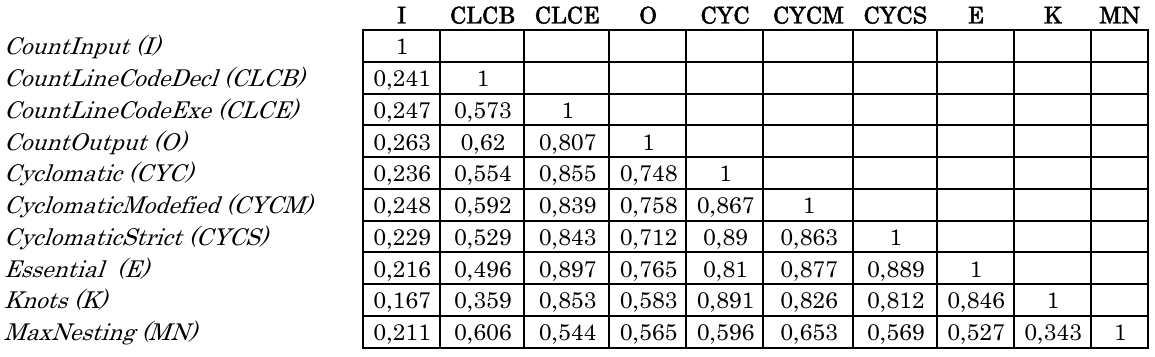}

\end{table}

\textcolor{black}{
Table~\ref{tab:UnderstandKendallCorrelation} presents the corresponding Kendall correlation results.
In this case, high correlations are limited to the cyclomatic variants only, with no additional metrics exceeding the cutoff.
This behavior is consistent with the more conservative nature of Kendall's $\tau$ and confirms that the strongest rank-based dependencies are confined within this metric family.
}

\textcolor{black}{Although Spearman and Kendall analyses revealed additional correlations for the Understand metrics, we retained the same metric set for subsequent analyses by selecting, among the three applied correlation filters (Pearson, Spearman, and Kendall), the most permissive one, namely Pearson's correlation, as the reference pruning criterion.
This choice was motivated by the intention to avoid an excessive reduction of the metric set and to preserve a broader range of candidate metrics.
Maintaining correlated metrics at this stage enables practitioners, in the final phase of the approach, to select those metrics to integrate into their quality assurance process based on their perceived interpretability and actionability, rather than relying exclusively on correlation strength.}

As a result, the final set of metrics for cross-project defect analysis remains unchanged and is reported in Table~\ref{tab:selected_metrics}.

\begin{table}[h]
\color{black}
\centering
\caption{Spearman rank correlation pairs ($|\rho| \geq 0.9$) for Understand metrics on the unified training dataset.}
\label{tab:UnderstandSpearmanCorrelation}
\small
\renewcommand{\arraystretch}{1.1}
\begin{tabular}{|l|l|c|}
\hline
\textbf{Metric 1} & \textbf{Metric 2} & $\boldsymbol{\rho}$ \\
\hline \hline
Cyclomatic & CyclomaticStrict & 0.990 \\ \hline
Cyclomatic & CyclomaticModified & 0.989 \\ \hline
CyclomaticModified & CyclomaticStrict & 0.980 \\ \hline
CyclomaticModified & MaxNesting & 0.956 \\ \hline
Cyclomatic & MaxNesting & 0.952 \\ \hline
CyclomaticStrict & MaxNesting & 0.945 \\ \hline
Cyclomatic & Knots & 0.905 \\ \hline
\end{tabular}
\end{table}

\begin{table}[h]
\color{black}
\centering
\caption{Kendall rank correlation pairs ($|\tau| \geq 0.9$) for Understand metrics on the unified training dataset.}
\small
\label{tab:UnderstandKendallCorrelation}
\renewcommand{\arraystretch}{1.1}
\begin{tabular}{|l|l|c|}
\hline
\textbf{Metric 1} & \textbf{Metric 2} & $\boldsymbol{\tau}$ \\
\hline \hline
Cyclomatic & CyclomaticStrict & 0.972 \\ \hline 
Cyclomatic & CyclomaticModified & 0.966 \\ \hline
CyclomaticModified & CyclomaticStrict & 0.944 \\ 
\hline
\end{tabular}
\end{table}

\subsubsection{Hypothesis Testing Analysis}

We used hypothesis testing to assess whether a given metric can discriminate between a faulty and a non-faulty function. Hypothesis testing is a statistical method used to determine whether there is sufficient evidence to support a specific hypothesis. For this purpose, we defined the following hypotheses:
\begin{enumerate}[label=$H_{\arabic*}$:]\setcounter{enumi}{-1}
    \item \textit{Faulty and non-faulty function samples belong to the same metric distribution.}
    \item  \textit{Faulty and non-faulty function samples do not belong to the same metric distribution.}
\end{enumerate}

To test these hypotheses, we used a non-parametric statistical test, specifically the Wilcoxon-Mann-Whitney test (WMW test), to evaluate $H_{0}$. We opted for a non-parametric test because, in a preliminary analysis, we verified that the sample distributions did not follow a normal distribution. The WMW test compares two independent samples to determine whether they originate from the same distribution, without assuming normality. The test produces a p-value, which determines whether $H_{0}$ can be rejected. A p-value below 0.05 allows rejection of $H_{0}$ , supporting the alternative hypothesis $H_{1}$.

\textcolor{black}{In addition to statistical significance, we also quantified the magnitude of the observed differences by computing Cliff's delta ($\delta$) for each metric. While p-values indicate whether a difference is statistically detectable, they do not convey the practical relevance of that difference, especially in large samples. Cliff's delta provides a non-parametric measure of effect size, capturing the degree of separation between faulty and non-faulty function distributions and thus complementing the hypothesis testing results with an interpretable measure of practical impact.}

The WMW test was applied to the metric sets obtained from both the Coverity and Understand tools for the overall dataset. The results, reported in Table~\ref{tab:wmw_cliff_coverity} and Table~\ref{tab:wmw_cliff_understand}, indicate that all metrics yielded p-values below 0.05. Consequently, we were able to reject $H_{0}$, confirming that faulty and non-faulty function samples do not belong to the same distribution.

\textcolor{black}{For the Coverity metrics, the corresponding Cliff's delta values are reported in Table~\ref{tab:wmw_cliff_coverity}. Most metrics exhibit medium to large effect sizes, indicating a substantial separation between faulty and non-faulty functions. In particular, \texttt{CCM}, \texttt{OperandCount}, and \texttt{OperationCount} show $\delta$ values close to or above 0.5, suggesting strong discriminative capability. \texttt{HalsteadEffort}, while statistically significant, presents a comparatively small effect size. Despite this, the metric was retained in the subsequent analyses to avoid overly restricting the candidate metric set. This decision reflects the intended use of the approach as a decision-support tool, where practitioners may prioritize interpretability and familiarity alongside statistical strength when selecting metrics for quality assurance activities.}

\begin{table}[h]
\color{black}
\centering
\small
\caption{Results of WMW test and Cliff's delta for Coverity metrics}
\label{tab:wmw_cliff_coverity}
\begin{tabular}{|l| c| c|}
\hline
\textbf{Metric Name} & \textbf{p-value} & \textbf{Cliff's $\lvert\delta\rvert$} \\
\hline \hline
CCM              & $3.12E{-109}$ & 0.530 \\ \hline
HalsteadEffort   & $8.71E{-108}$ & 0.093 \\ \hline
OperandCount     & $2.25E{-109}$ & 0.549 \\ \hline
OperationCount   & $2.39E{-90}$  & 0.497 \\ \hline
\end{tabular}
\end{table}

\textcolor{black}{
For the Understand metrics, the corresponding Cliff's delta values are reported in Table~\ref{tab:wmw_cliff_understand}.
A similar trend emerges, with all metrics showing statistically significant differences and predominantly medium-to-large effect sizes. Metrics capturing code size and structural complexity, such as \texttt{CountLineCodeExe}, \texttt{CountStmt}, and the \texttt{Cyclomatic} family, exhibit $\delta$ values consistently above 0.5, indicating strong discriminatory power. Other metrics present moderate effect sizes, which, while less pronounced, still contribute complementary perspectives on code structure and complexity. Maintaining this broader metric set supports a flexible, practitioner-oriented selection process in later phases of the approach.
}

\begin{table}[h]
\color{black}
\centering
\small
\caption{Results of WMW test and Cliff's delta for Understand metrics}
\label{tab:wmw_cliff_understand}
\begin{tabular}{|l| c| c|}
\hline
\textbf{Metric Name} & \textbf{p-value} & \textbf{Cliff's $\lvert\delta\rvert$} \\
\hline \hline
AltCountLineCode      & $1.38E{-121}$ & 0.591 \\ \hline
CountInput            & $2.06E{-45}$  & 0.354 \\ \hline
CountLine             & $5.19E{-121}$ & 0.597 \\ \hline
CountLineCodeDecl     & $2.33E{-65}$  & 0.395 \\ \hline
CountLineCodeExe      & $1.18E{-128}$ & 0.602 \\\hline
CountOutput           & $2.68E{-114}$ & 0.561 \\\hline
CountStmt             & $2.30E{-119}$ & 0.579 \\\hline
CountStmtDecl         & $1.96E{-66}$  & 0.403 \\\hline
CountStmtExe          & $2.90E{-124}$ & 0.590 \\\hline
Cyclomatic            & $1.56E{-111}$ & 0.542 \\\hline
CyclomaticModified    & $6.76E{-107}$ & 0.530 \\\hline
CyclomaticStrict      & $1.81E{-111}$ & 0.543 \\\hline
Essential             & $9.30E{-130}$ & 0.419 \\\hline
Knots                 & $5.85E{-99}$  & 0.490 \\ \hline
MaxNesting            & $1.30E{-96}$  & 0.496 \\\hline
\end{tabular}
\end{table}

The results of this hypothesis testing analysis demonstrate that the metrics provided by Coverity and Understand can effectively distinguish between faulty and non-faulty functions.

\subsection{Thresholds Definition}
\label{sec:thresholds}

Finally, in the \textit{Threshold Definition} step, we investigated whether each selected metric exhibited a threshold value that could statistically distinguish faulty functions from non-faulty ones with a meaningful level of confidence, using hypothesis testing. The goal of this phase was to determine whether faulty functions consistently exhibit a lower bound for each metric, such that values exceeding a given threshold indicate a high likelihood of fault-proneness.

\subsubsection{Faulty Function Thresholds Extraction}

To evaluate thresholds for faulty functions, the following null hypothesis $H_{0}$ and its alternative hypothesis $H_{1}$ were proposed:

\begin{enumerate}[label=$H_{\arabic*}$:]\setcounter{enumi}{-1}
    \item  \textit{For a given metric, do the faulty functions assume a value lower than or equal to $x$?}
    \item  \textit{For a given metric, do the faulty functions assume a value higher than $x$?}
\end{enumerate}

\textcolor{black}{
The one-sample Wilcoxon signed-rank test was applied to assess whether the location parameter of the metric distribution for faulty functions exceeded a given threshold.
}

\textcolor{black}{
The threshold extraction procedure can be formally interpreted as the inversion of a one-sample Wilcoxon signed-rank test.
Specifically, for each metric, we compute the one-sided $(1-\alpha)$ lower confidence bound of the faulty location parameter, corresponding to the Hodges--Lehmann pseudo-median, following standard results in nonparametric inference based on test inversion \cite{hollander2013nonparametric}.
This bound represents the smallest metric value such that, with confidence level $(1-\alpha)$, faulty functions exhibit values above it.
Under this interpretation, Type~I error is controlled at level $\alpha$ by construction, and no multiple-testing issue arises.}
\textcolor{black}{In practice, because all analyzed metrics assume discrete integer values, we retain an iterative scanning procedure for implementation convenience.
This procedure does not represent a sequence of hypothesis tests to be interpreted individually, but a numerical search for the lower confidence bound implied by the inverted Wilcoxon test.
The final threshold is rounded to the nearest admissible integer value.} A constant increment step of 1 was used in the algorithm because all the metrics under analysis, extracted from Coverity and Understand, assume values in the set of natural numbers.
This discrete, integer-based nature of the metric values justifies the use of a unitary step size without loss of granularity. \textcolor{black}{This integer rounding reflects the discrete nature of the metrics and does not affect the statistical interpretation of the confidence bound.}

In the following, we report the statistical procedure adopted to compute the thresholds, allowing for replication and adaptation in other industrial settings.
The pseudo-code of the implemented R script is shown in Algorithm~\ref{algo2}, summarizing the threshold computation procedure, and was applied independently to each metric collected from both analysis tools.
Due to confidentiality agreements with our industrial partner, we cannot disclose the exact threshold values derived from our analysis.

\begin{algorithm}[h!]
\SetAlgoLined
Threshold = 0\;
WilcoxonRankTest(Threshold, alternative = ``greater'')\;
\While{the null hypothesis is rejected}{
  Threshold = Threshold + 1\;
  WilcoxonRankTest(Threshold, alternative = ``greater'')\;
}
Threshold = Threshold - 1\;
\caption{\textcolor{black}{Numerical search for the one-sided lower confidence bound via Wilcoxon test inversion}}
\label{algo2}
\end{algorithm}

\section{Thresholds Validation}
\label{sec:empEvaluation}

In this section, we assess the effectiveness of the thresholds derived from our cross-project analysis.
The validation approach adopted in this study focuses exclusively on precision, as defined in Table~\ref{tab:measure}, rather than also including recall and accuracy, which are more typical in fault prediction evaluations. This decision was not arbitrary but guided by the specific quality assurance priorities expressed by our industrial partner.
During our collaboration, discussions with project managers and senior developers revealed that their primary interest was in assessing the ability of the proposed thresholds to minimize false positives, rather than in achieving exhaustive fault coverage. In their software development workflow, functions flagged as faulty trigger manual code inspections, documentation efforts, and compliance procedures.
As a result, precision (i.e., the proportion of flagged functions that are truly faulty) was considered the most relevant metric for evaluation. In contrast, recall (i.e., the proportion of actual faults successfully detected) and accuracy (i.e., the proportion of correctly classified instances overall) were considered secondary.
Prioritizing high precision helps reduce unnecessary quality assurance overhead and ensures that manual reviews are focused on components with the highest likelihood of containing real faults.

\begin{table}[H]
\caption{Evaluation measures. \\ Legend:  \\ True Negative (TN): Function with a metric value below Threshold \(T\) and not faulty. \\  True Positive (TP): Function with metric value above Threshold \(T\) and faulty. \\ False Negative (FN): Function with a metric value below Threshold \(T\) but faulty. \\ False Positive (FP): Function with a metric value above Threshold \(T\) but not faulty.}
\label{tab:measure}
\centering
\begin{tabular}{|l|c|}
    \hline
    \textbf{Measure} & \textbf{Formula} \\ \hline\hline
    Precision & $\frac{TP}{TP + FP}$ \rule{0pt}{2.5ex} \\ \hline
    Accuracy & $\frac{TP + TN}{TN + TP + FN + FP}$ \rule{0pt}{2.5ex} \\ \hline
    Recall & $\frac{TP}{TP + FN}$ \rule{0pt}{2.5ex} \\ \hline    
\end{tabular}

\end{table}

Based on this evaluation objective, we conducted a dedicated validation experiment to answer the following research question:
\begin{enumerate}[label=$RQ{\arabic*}$:] 
    \item \textit{What is the precision of code metric thresholds in identifying faulty functions?}
\end{enumerate}

To this end, we adopted a hold-out validation strategy to ensure a clear separation between the data used to derive thresholds and the data used to assess their effectiveness. Specifically, thresholds were computed using the datasets from Projects ${P_1}$ and ${P_2}$, whereas the dataset from Project ${P_3}$ was exclusively reserved for validation. Project ${P_3}$ was not used at any stage for metric selection, filtering, or threshold tuning; it served solely as an independent hold-out set for validation. Furthermore, rather than relying on balanced random sampling techniques that artificially equalize class distributions, the validation was conducted using the complete set of faulty and non-faulty functions from Project ${P_3}$. This decision was driven by the need to assess thresholds under realistic class imbalance conditions, reflecting the natural distribution of faults typically found in industrial codebases. Consequently, this evaluation strategy aims to provide a practical assessment of the thresholds' generalizability across projects, reflecting their effectiveness in real-world industrial applications and supporting actionable quality assurance processes.

Table~\ref{tab:precision_all} presents the evaluation results of the derived metric thresholds, reporting the precision values achieved for each selected metric from the \textit{Coverity} and \textit{Understand} static analysis tools. Overall, the results confirm the effectiveness of the defined thresholds in identifying faulty functions with high confidence. Among the Coverity-derived metrics, \texttt{CCM} achieved the highest precision (0.871), followed closely by \texttt{HalsteadEffort} (0.843) and \texttt{OperandCount} (0.841). These values indicate that the thresholds defined for these metrics can reliably flag faulty functions with minimal false positives, aligning with the industrial requirement for high-precision detection.

Similarly, metrics derived from Understand also demonstrated strong performance. The highest precision was observed for \texttt{CyclomaticStrict} (0.899), \texttt{Cyclomatic} (0.892), and \texttt{Knots} (0.891), underscoring the effectiveness of structural complexity indicators in identifying fault-prone code. Several other metrics, such as \texttt{Essential}, \texttt{CountLineCodeExe}, and \texttt{CyclomaticModified}, also achieved precision above 0.888, suggesting that control-flow complexity and code size remain valuable predictors in cross-project fault prediction. These results validate the practical relevance of the derived thresholds and demonstrate their potential for integration into software quality assurance pipelines, especially in safety-critical development environments where early, high-confidence fault detection is essential.

In the deployed workflow, thresholds are not combined into a single classifier. Each metric is applied independently to deliver actionable, per-metric feedback (e.g., if \texttt{CyclomaticStrict} is above threshold, reviewers focus on control-flow simplification). This preserves traceability, supports ISO 26262 audits, and enables targeted effort allocation. Aggregate rules (e.g., OR or k-of-m) would collapse distinct remediation cues into a single flag and thus reduce actionability.

\begin{table}[t]
\color{black}
\centering
\small
\caption{Coverity and Understand Metric Thresholds Evaluation}
\label{tab:precision_all}

\begin{minipage}{\linewidth}
\centering
(a) Coverity Thresholds Evaluation
\vspace{0.3em}

\begin{tabular}{|l|c|c|c|}
\hline
\textbf{Metric} & \textbf{Precision} & \textbf{Recall} & \textbf{Accuracy} \\
\hline \hline
CCM              & 0.871 & 0.453 & 0.680 \\  \hline
HalstedEffort    & 0.843 & 0.424 & 0.671 \\ \hline
OperandCount     & 0.841 & 0.421 & 0.682 \\ \hline
OperationCount   & 0.795 & 0.482 & 0.679 \\
\hline
\end{tabular}
\end{minipage}

\vspace{0.8em}

\begin{minipage}{\linewidth}
\small
\centering
{(b) Understand Thresholds Evaluation}

\vspace{0.3em}

\begin{tabular}{|l|c|c|c|}
\hline
\textbf{Metric} & \textbf{Precision} & \textbf{Recall} & \textbf{Accuracy} \\
\hline \hline
CountInput            & 0.751 & 0.440 & 0.646 \\ \hline
CountLineCodeDecl     & 0.790 & 0.438 & 0.660 \\ \hline
CountLineCodeExe      & 0.888 & 0.436 & 0.690 \\\hline
CountOutput           & 0.855 & 0.409 & 0.669 \\\hline
Cyclomatic            & 0.892 & 0.437 & 0.692 \\\hline
CyclomaticModified    & 0.881 & 0.413 & 0.678 \\\hline
CyclomaticStrict      & 0.899 & 0.443 & 0.689 \\\hline
Essential             & 0.888 & 0.379 & 0.665 \\\hline
Knots                 & 0.891 & 0.395 & 0.673 \\\hline
MaxNesting            & 0.856 & 0.328 & 0.635 \\
\hline
\end{tabular}
\end{minipage}

\end{table}


\color{black}
\section{Further Discussion} 
\label{sec:furdiscussion}

In this section, we further discuss the main results of our study. We analyze the
effectiveness of the derived metric thresholds, \textcolor{black}{provide an
aggregate assessment of recall and accuracy on the independent ${P_3}$ validation
dataset,} discuss the recall and accuracy values through empirical data and
manual inspection of false negatives, and examine the practical integration of
the selected metrics into the company's QA workflow based on focus group
feedback.

\subsection{Effectiveness of the Derived Thresholds}
\color{black}
The results presented in Table~\ref{tab:precision_all} demonstrate that the derived thresholds are effective in identifying functions with a high likelihood of being faulty. The consistently high precision values across both Coverity and Understand metrics indicate that when a function exceeds the identified threshold, it is very likely to be genuinely fault-prone. These findings confirm that the cross-project fault prediction process proposed in this study is capable of producing context-specific thresholds that can be used for early identification of high-risk components. This capability is particularly valuable in industrial settings, as it reduces the incidence of false positives and, consequently, minimizes unnecessary debugging and manual code inspection efforts. Rather than aiming to define universally applicable thresholds, this work focuses on introducing a structured and interpretable process for deriving \textit{context-specific thresholds} that are tailored to the characteristics of a given industrial environment. By emphasizing this methodological perspective, the study addresses a practical challenge noted by Mori \textit{et al.}~\cite{mori2018}, who argue that the absence of domain-adapted thresholds has long been a barrier to the widespread adoption of software metrics in real-world development settings.

\subsection{Complementary Analysis of Recall and Accuracy on ${P_3}$}

\textcolor{black}{ In addition to precision, we analyzed recall and accuracy on the ${P_3}$ dataset, which was exclusively reserved as an independent hold-out set for validation. While precision is discussed per metric to reflect how thresholds are consumed in the industrial QA workflow, recall and accuracy are analyzed here at an aggregate level to provide an overall view of performance on unseen data. }

\textcolor{black}{
Specifically, we report macro-averaged precision, recall, and accuracy aggregated across all selected metrics, separately for Coverity and Understand. Macro-averaging summarizes the collective behavior of the retained metrics and complements the per-metric analysis by offering a tool-level perspective. To characterize variability in a compact and robust way, interquartile dispersion bands (Q1--Q3) are reported alongside the mean values.} 

\textcolor{black}{ Table~\ref{tab:macro_pra_p3} reports the resulting macro-averaged performance on the ${P_3}$ dataset.}

\begin{table}[h] 
\centering 
\small 
\setlength{\tabcolsep}{4pt} 
\renewcommand{\arraystretch}{1.2} \caption{Macro-averaged precision, recall, and accuracy on the ${P_3}$ dataset, aggregated across all selected metrics, with interquartile dispersion bands (Q1--Q3).} \label{tab:macro_pra_p3} 
\color{black} 
\begin{tabularx}{\textwidth}{lccc ccc ccc} 
\toprule 
\textbf{Tool} & \multicolumn{3}{c}{\textbf{Precision}} & \multicolumn{3}{c}{\textbf{Recall}} & \multicolumn{3}{c}{\textbf{Accuracy}} \\ \cmidrule(lr){2-4} \cmidrule(lr){5-7} \cmidrule(lr){8-10} & \textbf{Mean} & \textbf{Q1} & \textbf{Q3} & \textbf{Mean} & \textbf{Q1} & \textbf{Q3} & \textbf{Mean} & \textbf{Q1} & \textbf{Q3} \\ 
\midrule
    Coverity & 0.837 & 0.829 & 0.850 & 0.445 & 0.423 & 0.460 & 0.678 & 0.677 & 0.680 \\ 
    Understand & 0.859 & 0.855 & 0.890 & 0.411 & 0.398 & 0.437 & 0.669 & 0.661 & 0.686 \\ \bottomrule 
\end{tabularx} 
\end{table}
\textcolor{black}{As shown in Table~\ref{tab:macro_pra_p3}, macro-averaged precision is consistently high for both tools, whereas recall and accuracy exhibit more moderate values. This pattern is stable across Coverity and Understand and reflects the expected trade-off when prioritizing precision in a highly imbalanced validation setting.}

For Coverity metrics, the average recall was 0.445, and the average accuracy was 0.678. For Understand metrics, the corresponding averages were 0.411 for recall and 0.669 for accuracy. While these values are more moderate than the achieved precision scores, indicating that some faulty functions remain undetected, they reflect the natural limitations of relying solely on structural metrics for fault detection.

In particular, the thresholds proved highly effective in capturing faults linked to complexity and maintainability, while some types of defects, such as semantic or logic-related errors, may require complementary analysis techniques. Rather than indicating a weakness, these results suggest that our approach can serve as a reliable component within a broader fault prediction strategy that incorporates
multiple analysis dimensions.

To further interpret the observed recall and accuracy limitations, we performed a manual inspection on approximately 20\% of the false negatives (i.e., faulty functions that did not exceed any identified thresholds). This analysis revealed that many of these functions were not characterized by structural complexity, but instead contained logical or semantic defects. Examples include incorrect variable initialization or faulty condition checks that do not necessarily
increase the metric values derived from static analysis. These issues often occur in syntactically simple code and are not adequately captured by structural metrics such as code size or nesting depth. Consequently, such defects may go unnoticed by threshold-based methods focused on static complexity indicators.
This insight reinforces the notion that our approach is most effective at
detecting structurally complex or maintainability-related issues, while being less sensitive to subtle semantic faults.

Nonetheless, these limitations were considered acceptable within the industrial context of our study. During multiple focus group sessions with project managers and senior developers, it became clear that maximizing fault coverage was not the primary objective. Instead, the organization prioritized the reduction of false positives, which often leads to inefficient use of verification resources. In safety-critical domains such as automotive firmware, where standards such as ISO~26262 require rigorous documentation and formal reviews, code inspections are both resource-intensive and costly. From the company's perspective, it is more valuable to identify a smaller subset of truly fault-prone functions, ensuring focused and justifiable quality assurance efforts, than to expand detection at the cost of increasing false alarms.

\subsection{Metric Selection for QA Integration}
\color{black}
To evaluate the practical relevance of the derived thresholds, we conducted a focus group with the project manager and senior developers from the company. The objective was to identify which of the evaluated metrics could be realistically integrated into their development workflow. Together, we agreed to retain only those metrics that achieved a minimum precision of 0.85. 
As a result, \texttt{Cyclomatic Complexity (CCM)} was selected as the sole Coverity metric to be used in the company's fault prediction process, as it met the precision requirement and was considered easy to interpret and actionable. For the Understand tool, the retained metrics included \texttt{CountLineCodeExe}, \texttt{CyclomaticModified}, \texttt{CyclomaticStrict}, \texttt{CountOutput}, and \texttt{MaxNesting}. On the other hand, metrics such as \texttt{Essential} and \texttt{Knots}, despite having high precision, were excluded due to concerns about their interpretability and limited practical utility in guiding fault-reduction activities. This selection process emphasizes the importance of balancing empirical performance with developer usability. Metrics that are not only precise but also meaningful and understandable to practitioners are more likely to be adopted and used consistently across teams.

Moreover, the per-metric design was chosen to maximize actionability: each violation maps to a specific remediation (reduce complexity, split functions, limit nesting, etc.), enabling teams to optimize developer hours on code changes with the highest expected payoff. This rationale is also why we did not adopt ML-based fault prediction: despite potential gains in aggregate accuracy, ML models typically provide non-prescriptive, less localized signals, which are harder to translate into concrete refactoring steps within a safety-critical process.

\section{Threats to validity}
\label{sec:Threats}

The validity of our empirical results is subject to several threats, categorized into construct validity, internal validity, external validity, and reliability. We discuss each threat explicitly and propose mitigation strategies where feasible.

\paragraph{\textbf{Construct Validity}}

A key threat to construct validity in this study concerns the representativeness and appropriateness of the selected software metrics and their derived thresholds for distinguishing between faulty and non-faulty functions. The metrics used were extracted from two static analysis tools, Coverity and Understand, both widely adopted in industry. While these tools provide a comprehensive set of well-established metrics, the initial selection was constrained by tool availability and relevance to the industrial partner's development practices, rather than derived from an exhaustive theoretical framework. This introduces the risk of omitting potentially informative metrics that are either unavailable in these tools or not prioritized in the industrial workflow.

Additionally, the process of empirically deriving metric thresholds, though grounded in rigorous statistical hypothesis testing, inevitably involves context-specific decisions that may introduce subjectivity. In particular, the selection of a minimum precision threshold (set to 0.85) as a filtering criterion for retaining metrics was made in collaboration with senior developers and the project manager during a focus group. While this decision aligns with industrial priorities (favoring fewer false positives), it also introduces a form of construct bias. Metrics that may offer high recall or balanced performance but slightly lower precision were excluded, potentially narrowing the scope of the model to only those metrics that favor conservative fault prediction behavior. This decision reflects a practical trade-off, emphasizing actionable and trustworthy predictions at the expense of broader metric diversity.

To mitigate these threats, we adopted a multi-step filtering process combining statistical tests (e.g., Wilcoxon signed-rank test), redundancy removal through correlation analysis, and expert validation. The selected metrics were thus not only statistically discriminative but also aligned with real-world interpretability and applicability in the industrial context.

\paragraph{\textbf{Internal Validity}} A significant threat to internal validity concerns potential biases introduced during the data collection and threshold-setting processes. The identification of faulty functions relied primarily on Jira issue reports, which introduces risks related to inaccurate or incomplete fault reporting, such as incorrectly logged issues or unreported faults. Additionally, our decision to measure software metrics exclusively from code versions immediately preceding the creation of Jira issues assumes that faults arose precisely at these points, potentially overlooking pre-existing faults or subsequent code modifications.

To mitigate these concerns, we established a rigorous and systematic procedure linking each Jira issue explicitly to specific code changes, confirmed through associated pull requests and manual verification. Nevertheless, the inherent limitations of relying solely on Jira-based issue tracking present a residual threat that may not be fully addressed without supplementary fault validation approaches. Another potential threat concerns the threshold used to filter highly correlated metrics during the metric selection phase. We adopted a conservative cutoff of 0.9 to retain metrics that, while moderately correlated, may still provide complementary and practically meaningful insights in an industrial context. However, this choice does not guarantee full statistical independence among the selected metrics. It is possible that metrics with correlations in the 0.5–0.9 range share redundant information, which could affect the precision or generalizability of the derived thresholds. Future replications could explore alternative cutoff values and apply dimensionality reduction techniques to assess the robustness of the selected metric set.

\textcolor{black}{An additional threat to internal validity concerns the stability of the derived thresholds under alternative design choices. In our setting, the thresholding outcome depends on multiple interacting design dimensions, including temporal sampling, project selection, metric filtering, and statistical configuration. As a result, defining a unique and well-founded notion of relative delta would require introducing arbitrary perturbation assumptions, potentially leading to misleading interpretations. For this reason, and to preserve methodological clarity, we did not include a global stability analysis based on relative deltas in the present study. As future work, this limitation could be addressed by isolating and perturbing individual design dimensions of the pipeline one at a time (e.g., temporal granularity or metric selection), enabling a more interpretable and dimension-specific stability assessment.}

\textcolor{black}{Moreover,} future replications of the process could explore single-step alternatives such as VARL or ROC-Youden thresholding combined with formal multiple-testing control.

\paragraph{\textbf{External Validity}} The external validity of our study may be limited by its specific industrial and technological context—embedded firmware developed in the C programming language for automotive applications under ISO 26262 compliance. Consequently, the selected software metrics, derived thresholds, and predictive outcomes are inherently tailored to this environment and may not generalize directly to other embedded software domains, different programming languages, or alternate development methodologies. Additionally, our dependence on the adopted static analysis tools further restricts generalizability, as the metric definitions, computation methods, and accuracy of these tools might differ from alternative analysis tools available in other contexts. To address these limitations and support broader applicability, we have provided detailed documentation of our statistical approach and threshold derivation methods, thus facilitating replication, validation, and potential adaptation of our methodology in analogous industrial contexts.
Furthermore, another potential threat concerns the limited number of projects (three) included in this study. Despite this limitation, the selected projects were not chosen arbitrarily. In collaboration with our industrial partner, we deliberately selected three firmware projects that share similar architectural features, development processes, and safety-critical requirements. This alignment ensured that the cross-project fault prediction scenario we evaluated was both realistic and relevant within the company's operational context. While this may limit generalizability to other types of systems, the consistency across projects reinforces the validity of the observed patterns and provides a solid foundation for demonstrating the applicability of the proposed methodology. Expanding the study to additional projects in the future will be instrumental in further validating the generalizability of our findings.

This study did not include an explicit comparison with published thresholding techniques such as percentile-based cutoffs, VARL, or ROC-Youden. This limitation results from confidentiality constraints that prevent disclosure of derived or comparative threshold values on the industrial datasets. However, the statistical logic of our approach, non-parametric hypothesis testing for distributional separation, shares conceptual similarities with these baseline methods. Future replications using open or anonymized datasets could reproduce our process alongside such baselines to quantify performance differences in terms of precision, recall, and accuracy.

\paragraph{\textbf{Reliability}} A significant threat to reliability pertains to the repeatability of the metrics extraction and threshold derivation processes. Although static analysis tools typically produce consistent results in extracting metrics, the iterative statistical procedures employed for threshold determination, particularly the use of automated R scripts and statistical tests, might introduce variability when executed repeatedly or applied to differing or extended datasets. Such variations could potentially impact the reproducibility of threshold values. 

To mitigate this risk, we have comprehensively documented the statistical methodology and provided detailed pseudocode outlining the threshold derivation algorithms, thereby enabling external researchers and practitioners to independently replicate and verify the process with greater precision.

\section{Conclusions and Future Works}
\label{sec:ConcFut}

In the following, we present the final considerations of our study. First, we summarize the main contributions and findings of our proposed process for deriving context-specific metric thresholds in industrial settings. Then, we outline several future research directions aimed at extending the applicability, generalizability, and technological integration of our approach.

\subsection{Conclusions}
\color{black}
In this work, we presented a structured process designed to address the challenge of defining software metric thresholds suitable for integration into fault detection workflows in industrial settings. Although machine learning and deep learning approaches often achieve high predictive accuracy, their inherent lack of transparency and interpretability limits their adoption in industrial settings. Industries, particularly those operating in safety-critical domains, require clear, actionable feedback to support compliance with evidence-based software quality standards such as ISO 26262. Statistical threshold-based methods effectively address this need by providing developers with transparent, actionable insights on code improvements.

The proposed process was developed and validated through an industrial collaboration and is composed of three main phases. First, we constructed function-level datasets by extracting and organizing software metrics and fault labels from three real-world firmware projects. This step relied on a dedicated traceability model and the use of static analysis tools (Coverity and Understand) to measure a wide set of product metrics. Second, we carried out a metric selection phase that combined visual inspection, correlation filtering, and focus group sessions with domain experts to retain only the most interpretable and statistically discriminative metrics. Third, we applied non-parametric statistical hypothesis testing to derive threshold values for the selected metrics. 

Our contribution lies not in proposing universally optimal thresholds, but in introducing a practical and replicable process for deriving context-specific thresholds that can be realistically adopted within industrial workflows. By leveraging the metrics already collected in the CI/CD pipeline, our approach enables early fault detection without requiring changes to the existing toolchain, thus facilitating adoption.
To evaluate the derived thresholds, we focused on precision, reflecting the industrial partner's priority of minimizing false positives. Since each flagged function requires costly manual review and documentation, high-confidence predictions were favored over broader fault coverage. 
The evaluation, conducted on a third independent project confirmed that several of the derived thresholds achieve high precision, making them well-suited for integration into quality assurance activities.
These results demonstrate that statistical thresholds can deliver precise early warnings of fault-prone functions, enabling developers to take corrective actions before defects propagate into later stages of development, a benefit particularly critical in safety-critical domains, where last-minute fixes may severely compromise traceability and regulatory compliance.

\subsection{Future Works}
\color{black}
Several future research directions will build upon and extend our current results. First, we plan to classify the types of faults present in the analyzed projects to evaluate whether specific error categories achieve higher recall rates. This analysis will help determine if certain defect types are more effectively detected by the defined metric thresholds. Gaining a deeper understanding of the relationship between fault types and metric-based thresholds could enhance the applicability of our approach, enabling more targeted and accurate fault prediction models.

To mitigate the moderate levels of recall and accuracy observed during validation, we plan to extend our approach by combining metric-based thresholds with complementary strategies capable of capturing faults not linked to structural complexity. In particular, we aim to explore hybrid fault prediction models that combine threshold-based methods with interpretable machine learning techniques. This direction seeks to enhance fault detection coverage while preserving transparency.

An additional future direction could be the exploration of using LLMs to support the early detection of faulty functions due to semantic or logic-related defects, which often go undetected by structural metrics. In this context, we are interested in investigating how LLMs can assist developers in interpreting threshold violations, suggesting context-aware refactorings, or offering human-readable explanations that align with auditability and safety standards. In particular, we aim to experiment with LLM-based static analysis workflows where natural language prompts are used to query code for potential logical inconsistencies, anti-patterns, or incomplete implementations. This could enable the early identification of faults that do not necessarily correlate with metric anomalies.

Moreover, we plan to compare the thresholds derived in this study with alternative thresholds computed using methodologies from the literature, providing a benchmark for our approach. Additionally, we plan to evaluate the stability of the derived thresholds over time by applying them to future versions of the same software products. A longitudinal analysis will help determine whether thresholds require recalibration as codebases evolve and whether patterns of fault-proneness shift due to architectural or process changes. 

Future work could also include a baseline comparison with standard threshold derivation methods (percentile, VARL, ROC-Youden) on public or sanitized datasets to strengthen external validity.

Finally, we plan to further investigate the effectiveness of these thresholds in an industrial setting by integrating them into the software quality assurance process. This validation will involve collaboration with developers working on active projects, providing real-world insights into the practical value of the thresholds in improving software quality. To support this step, we also aim to incorporate thresholds into quality dashboards and CI/CD workflows, enabling actionable alerts and early feedback. We will evaluate the usability and acceptance of these mechanisms through workshops and focus groups with practitioners, aligning future improvements with concrete industrial needs.

\bibliographystyle{elsarticle-num} 
\bibliography{FaultBiblio}
\end{document}